\documentclass[10pt,emptycopyrightspace]{ewsn-proc}

\renewcommand{\paragraph}[1]{\vskip 6pt\noindent\textbf{#1 }}

\usepackage{graphicx}
\usepackage{balance}
\usepackage{comment}
\usepackage{amsmath, amssymb,algorithm,epstopdf,url}
\usepackage{algpseudocode,subfigure}
\usepackage{color}

\newcommand{\COMMENTED}[1]{}

\numberofauthors{1}

\graphicspath{{Pictures/}}
\newtheorem{theorem}{Theorem}
\newdef{definition}{Definition}

\author{
\alignauthor Seyed-Mohsen Moosavi-Dezfooli$^1$, Yvonne-Anne Pignolet$^2$, 
Dacfey Dzung$^2$ \\
        \affaddr{$^1$\'Ecole Polytechnique F\'ed\'erale de Lausanne, 
				Switzerland, $^2$ ABB Corporate Research, Switzerland}\\
       \email{seyed.moosavi@epfl.ch, yvonne-anne.pignolet@ch.abb.com, 
			dacfey.dzung@ch.abb.com}
}

\title{Simultaneous Acoustic Localization of Multiple Smartphones with 
Euclidean Distance Matrices}

\CopyrightYear{2015}

\begin{document}

\maketitle
\begin{abstract}
In this paper, we present an acoustic localization system for multiple devices. 
In contrast to systems which localize a device relative to one or several anchor
points, we focus on the joint localization of several devices relative to each 
other.
We present a prototype of our system on off-the-shelf smartphones.
No user interaction is required, the phones emit acoustic pulses
according to a precomputed schedule. Using the elapsed time between two times of
arrivals (ETOA) method with sample counting, distances between the devices 
are estimated. These, possibly incomplete, distances are the input to an 
efficient and robust multi-dimensional scaling algorithm returning a position 
for each phone.
We evaluated our system in real-world scenarios, achieving error margins 
of 15 cm in an office environment.\footnote{This is the extended version of a paper accepted at International Conference on Embedded Wireless Systems and Networks (EWSN) 2016.}
%

\end{abstract}

\category{C.2.1}{Computer-communication networks}{Network Architecture and Design}[Wireless Communication]

\terms{Algorithms, Measurements}

\vspace{.1cm}
\noindent\emph{Keywords:} {indoor localization, acoustics, EDM, ETOA}

\section{Introduction}
  \label{sec:intro}
\textbf{Motivation}
Smartphones and other mobile devices have become ubiquitous in our lives. 
As a consequence, a large variety of location-dependent applications emerge to
support users at work, in shopping malls, airports, railway stations, museums and 
exhibitions. While GPS provides localization outdoors, it is often not useable
inside or the localization it provides is too coarse. Thus, a multitude of 
localization approaches based on Wi-Fi or sensors or smartphones have been 
devised.

In this paper we address localization using acoustic signals, as every 
commercial-off-the-shelf (COTS) smartphone is equipped with a microphone and 
speaker. In particular, we focus on the localization of several devices relative
to each other. Such a system can be used for e.g., asset tracking, to ensure 
safety around (unmanned) vehicles and machines in industrial settings, for 
augmented reality applications, either alone or complementing other 
localization systems.

\textbf{From Ranging to Positioning}
To find the distance between two devices without a measuring tape, there are 
many options. E.g., electromagnetic or audio waves can be used to estimate
the distance based on the time they need to propagate from one device to the 
other. Methods based on accelerometer measurements (pedestrian dead reckoning), 
or pictures taken by cameras can be used to determine
distances. But once distances are known, how can you find the position of a
specific node with respect to some coordinate system? If there are nodes of
which we know the positions, a ranging method helps us to find the distance 
between these nodes and the unknown one. Using these distances, there are 
several ways, e.g., trilateration to find the position of the unknown node. On 
the other hand if we know that two other nodes are close, and we know the 
location of one of them, we can say with some amount of confidence that the 
other one has more or less the same location and use nearest neighbor methods 
like \cite{cover1967} to determine positions of unknown nodes.

Starting from a simple acoustic ranging application, we propose methods and 
algorithms to enable the calculation of the position of several phones 
simultaneously. To this end we (i) design a pulse shaping and detection scheme 
which has not been used for acoustic ranging to the best of our knowledge and 
(ii) we propose an algorithm to schedule the actions of recording, emitting a 
pulse and stopping on the phones and (iii) solve the resulting multi-dimensional
scaling problem with Euclidean distance matrices (EDMs).

Using COTS smartphones entails a set of disadvantages and constraints. Both the 
speaker and microphone systems are optimized for voice, i.e., they are designed 
for signals in the range of 20 Hz to 22 kHz, higher frequencies are affected by 
lowpass filters in the audio chain of smartphones \cite{filonenko2013}.
Moreover, the API to access the audio system of a phone is limited, so 
using it for ranging and positioning algorithms 
is not straight-forward. We describe how we cope with these constraints when 
building a prototype on Samsung S4 mini phones and we validate our system in 
real-world scenarios reflecting the conditions of an office environment.

\textbf{Contributions: }
We demonstrate in this paper how acoustics can be used to
calculate positions of several devices relative to each other without
anchor nodes. Compared with other systems, some of which rely on specialized
hardware, our system features a low-cost deployment as well as accuracy. 
We use BeepBeep~\cite{peng2007} ranging method as a basis, 
overcoming difficulties due to the multi-phone settings with a different pulse 
and detection scheme and proposing a scheduling scheme to deal with collisions 
to build a reliable system.We reduced the abstract problem behind to MDS and 
designed a novel weighting scheme dealing with possibly large (non-Gaussian) 
ranging errors.
Hence, our system and validation work feature the following.
 \begin{itemize}
\item Protocol and algorithm for relative positioning of several devices at the same time. 
\item Robustness. It is not necessary that each device determines the distance
to all other devices. Incomplete distance matrices suffice for localization.
\item No anchor points or synchronization. Even if clocks are not 
synchronized and no anchor positions are known, our system can localize devices.
\item Evaluation. We implement our system on Android smartphones and evaluate it
in office environment scenarios. The mean location error is 5-15 centimeters
 depending on the environment and configuration,
satisfying the requirement of many applications.
\end{itemize}

The rest of the paper is organized as follows. Section~\ref{sec:background} 
presents background knowledge on ranging and pulse shaping methods as well as 
multi-dimensional scaling. Then a simple acoustic ranging application and its 
implementation are described in Section~\ref{sec:ranging}. The design of our multi
-node localization system design is discussed in Section~\ref{sec:multi} and 
evaluated in Section~\ref{sec:eval}. In Section~\ref{sec:relwork}, we 
review related work followed by a discussion and conclusion in 
Section~\ref{sec:conclusion}.

\section{Background}\label{sec:background}
\subsection{Acoustic Ranging}
In this section, we show how to use audio technology, i.e. microphones and 
speakers, in order to implement a meter for smartphones, which measures the 
distance between two phones. We can use the characteristics of sound waves to 
determine the traveled distance. There are various methods to do so but broadly 
speaking we can distinguish between two types, time-based and power-based 
methods. We summarize the most used time-based ones here.

The propagation time of waves between two nodes is a measure that tells us 
information about the distance between them. If we have information about speed 
of a wave in an environment, the propagation time can be translated into 
distance. For the audio waves, the propagation speed depends on the properties 
of the substance through which the wave is traveling. A practical model for the 
propagation of sound wave through dry air is as follows
\begin{equation}
\nu_s=331.3+0.606\theta\quad\frac{m}{s}.
\end{equation}
This simplified linear model only depends on the temperature $\theta$ in degrees
Celsius of the environment. Given this model and the propagation time 
$\Delta_p$, one can calculate the distance between two nodes,
\begin{equation}
d=\nu_s\Delta_p.
\end{equation}
Using smartphones, finding $\Delta_p$ is not as easy as it sounds. There are 
various methods and tactics to do so. The simplest thing to do is to calculate $
\Delta_p$ by using the arrival and departure time of a sound signal. Suppose 
phone 1 sends a sound pulse (we well discuss it in detail later) at time $T_1$ 
and phone 2 receives (records) it at time $T_2$ then $\Delta_p=T_2-T_1$.
This is the base of Time-of-Arrival (TOA) methods. One important aspect of 
measuring a time difference is to have the clocks synchronized to determine
$T_1$ and $T_2$. In theory, the synchronization can be done through a faster 
signal than sound like radio signals. Unfortunately, commercial off-the-shelf 
phones do not offer tight time synchronization and the Android environment does 
not allow to schedule the execution of tasks at a high time resolution.

To avoid time synchronization, other methods can be used. They can provide 
$\Delta_p$ indirectly. The round trip time (RTT) is a quantity that does not 
require time synchronization. Suppose phone 1 sends a sound pulse at time $T_1$.
As soon as phone 2 receives phone 1's pulse, it sends another pulse which will 
be received at phone 1 at time $T_1'$. Then $\Delta_p=(T_1'-T_1)/2$.
If there is no additional delay between receiving the first pulse and sending 
the second one, this method works fine. Since $T_1$ and $T_1'$ are measured on 
the same device, there is no need for time synchronization. However this method 
is not feasible on an Android smartphone. There are many OS mechanisms such as
garbage collection that introduce delays in the procedure between the time that 
phone 2 receives phone 1's pulse and the time that it sends its own pulse. 

Hence, the following scheme can be used to calculate $\Delta_p$:
\begin{itemize}
\item Phone 1: Generate a pulse to send over microphone, store timestamp of this
event as $T_1$
\item Phone 2: Listen and detect the pulse sent by Phone 1, store timestamp of 
this event as $T_2$
\item Phone 2: Generate a pulse to send over microphone, store timestamp of this
event as $T_2'$
\item Phone 1: Listen and detect the pulse sent by Phone 2, store timestamp of 
this event as $T_1'$
\end{itemize}

Now the time traveled by the sound wave between two phones can be calculated as
\begin{equation}
\Delta_p=\frac{(T_1'-T_1)-(T_2'-T_2)}{2}
\label{equ:rtt_delay}
\end{equation}
In (\ref{equ:rtt_delay}), we subtract the mentioned delay, e.g. OS delay, 
$T_2'-T_2$ to get only the duration that sound waves were on the fly. This 
method is called elapsed time between the two time-of-arrivals (ETOA) which is 
proposed in \cite{peng2007} for acoustic ranging.

\subsection{Pulse Shaping}
Any time measuring method needs an accurate pulse detection scheme. The optimal detector in the sense of Signal-to-Noise Ratio (SNR) is a 
matched filter \cite{haykin2013}.

Suppose we send a pulse called $s(t)$. On the receiver side we have
\begin{equation}
r(t) = s(t-\Delta) + w(t),
\end{equation}
where $w(t)$ is a signal which is independent of $s(t)$. To find $\Delta$, we 
pass $r(t)$ through a matched filter, i.e. $h_{MF}(t)=s(-t)$, hence 
\begin{equation}
y(t)=r(t)*h_{MF}(t)=s(t-\Delta)*s(-t)+w(t)*s(-t).
\label{equ:matched_filter}
\end{equation}
Since $w(t)$ and $s(t)$ are independent, for $\Delta$ we have
\begin{equation}
\Delta=\text{arg}\max_t y(t).
\label{equ:delta}
\end{equation}
In an ideal world, the equality in (\ref{equ:delta}) holds. However, in the real
world it might not hold because the second term in the rhs of 
(\ref{equ:matched_filter}) is no longer zero. Therefore, we need a pulse shape 
with a narrow autocorrelation function. It means that a pulse with a higher 
bandwidth makes detection easier and more accurate. However, the available 
bandwidth is limited on smartphones because of the frequency response of their 
microphone and speaker. Hence, we have to compromise between the bandwidth and 
the 
detectability.

A very simple option for a pulse shape is a finite duration sinusoidal signal, 
although it has a low bandwidth and a low detectability. In the presence of 
noise and interference, the accuracy of detection with pure sinusoids drops. 
Another candidate is a chirp signal, a frequency variant sinusoid, 
used in ~\cite{peng2007} and ~\cite{hoflinger2012} for acoustic ranging. 
Pseudo-random sequences have been widely used in wireless communications
contexts \cite{vucetic1997} due to their narrow autocorrelation. PN sequences 
are almost white noise but they differ in the distribution. As we will 
explain in more detail later, we use pseudo-random sequences for our setting 
because they enable an easier implementation for multi-user detection and 
have a narrower autocorrelation than the other variants.


\subsection{Multi-Dimensional Scaling}\label{sec:EDM}
Given two phones and their pairwise distance, we cannot determine their 
corresponding locations. Even if we have the location of one of the nodes, 
the ambiguity in the location of the other one remains at a circle around the 
known one unless there is more information available. If we have the distances 
between these two phones and another phone with a known location, the ambiguity 
decreases to just two points. A fourth phone with known location can resolve any
ambiguity. This is the principle of trilaterization. 
Even if all the locations are unknown, we can find the location of the phones up
to an affine transform (rotation and translation) thanks to the mathematical 
tool called Euclidean Distance Matrix (EDM).

First, we review some basics of EDM and its properties. Next, we show how we can
use EDM as a tool to find the location of devices.

\subsubsection{Euclidean Distance Matrix (EDM)}
Consider a list of points $\{\mathbf{x}_1,\mathbf{x}_2,\dots,\mathbf{x}_N\}$ in 
the Euclidean space $\mathbb{R}^\eta$ of dimension $\eta$. An Euclidean Distance
Matrix (EDM) is a matrix $\mathbf{D}$ such that
\begin{equation}
\mathbf{D}[i,j]=d_{i,j}^2=\|\mathbf{x}_i-\mathbf{x}_j\|^2.
\end{equation}
In other words, each entry of $\mathbf{D}$ is an Euclidean distance-square 
between pairs of $\mathbf{x}_i$ and $\mathbf{x}_j$. Due to the Euclidean metric
properties, the elements of $\mathbf{D}$ satisfy the following.
\begin{enumerate}
\item Non-negativity: $d_{i,j}\geq 0$ for all $i,j$.
\item Self-distance: $d_{i,j}=0\Leftrightarrow \mathbf{x}_i=\mathbf{x}_j$.
\item Symmetry: $d_{i,j}=d_{j,i}$ for all $i,j$.
\item Triangle inequality: $d_{i,j}\leq d_{i,k}+d_{k,j}$ for all $i,j,k$.
\end{enumerate}
While every EDM satisfies these properties, they are not sufficient conditions 
to form an EDM. We bring a theorem from \cite{schoenberg1935} that states the 
necessary and sufficient conditions for a matrix to be an EDM but let us give 
some definitions first.
\begin{definition}[Symmetric hollow subspace]
Denoted by $\mathbb{S}^N_h$, the symmetric hollow space is a proper subspace of 
symmetric matrices $\mathbb{S}^N$ with a zero diagonal.
\end{definition}
\begin{definition}[Positive semi-definite cone]
Denoted by $\mathbb{S}^N_+$, the positive semi-definite cone is the set of all 
symmetric positive semi-definite matrices of dimension $N\times N$.
\end{definition}
\begin{definition}
The geometric centering matrix $\mathbf{L}$ is defined as
\begin{equation}
\mathbf{L}\triangleq\mathbf{I}-\frac{1}{N}\mathbf{1}\mathbf{1}^T
\end{equation}
where $\mathbf{I}$ is the $N\times N$ identity matrix and $\mathbf{1}$ is the 
all one column vector in $\mathbb{R}^N$.
\end{definition}
The necessary and sufficient conditions for an $N\times N$ matrix $\mathbf{D}$ 
to be an EDM are
\begin{theorem}[{Schoenberg \cite{schoenberg1935}}]
\begin{equation}
\text{\textbf{D} is an EDM}\Leftrightarrow
\left\{\begin{array}{l l}
-\mathbf{LDL}\in\mathbb{S}^N_+
\\\mathbf{D}\in\mathbb{S}^N_h
\end{array}\right.
\end{equation}
\end{theorem}
\begin{theorem}
Assume $\mathbf{T}$ is an isometric transformations. We have,
\begin{equation}
\mathbf{D}(\mathbf{T}(\mathbf{X}))=\mathbf{D}(\mathbf{X}).
\end{equation}
\label{thm:rigid_transform}
\end{theorem}

\begin{algorithm}[t]
\caption{Classical MDS \cite{shang2003}}
\textbf{Input:} Dimension $\eta$, estimated squared EDM $\mathbf{D}$\\
\textbf{Output:} Estimated positions
\begin{algorithmic}[1]
\State Compute $(-1/2)\mathbf{LDL}$;
\State Compute the best rank-$\eta$ approximation $\mathbf{U}_\eta\Sigma_\eta 
\mathbf{U}_\eta^T$ of $(-1/2)\mathbf{LDL}$ using Singular Value Decomposition
(SVD)
\State Return $\mathbf{U}_\eta\Sigma_\eta^{1/2}$ as the estimated positions
\end{algorithmic}
\label{alg:cmds}
\end{algorithm}

Suppose a situation where the location of the nodes, i.e. 
$\{\mathbf{x}_1,\mathbf{x}_2,\dots,\mathbf{x}_N\}$, are unknown but their 
corresponding EDM is given. The goal is to find the set 
$\{\mathbf{x}_1,\mathbf{x}_2,\dots,\mathbf{x}_N\}$ based on the given EDM. As 
Theorem \ref{thm:rigid_transform} proposes, the solution is not unique but for 
our purpose, one of the possible solutions is still desirable.

Many methods to solve this problem are proposed in the literature. The classical
approach to solve this problem is called classical Multi-Dimensional Scaling 
(cMDS), originally proposed in psychometrics \cite{kruskal1964}.
In an error-free setup where the all the pairwise distances are measured without
error, cMDS exactly recovers the configuration of the points \cite{shang2003}. This method is simple and efficient. However, in a noisy situation it does not 
guarantee the optimality of the solution. Furthermore, it can only be used if 
all distances are known.

\section{Building Block: Acoustic Ranging App}
\label{sec:ranging}

We have discussed the basic ideas to implement an acoustic meter in the previous
section. To make a real application that measures distances, we need to tackle 
some constraints due to the chosen platform.

\subsection{Time Stamping and Sample Counting}

As mentioned earlier, the Android OS has many sources of unpredictable delays 
and synchronization of phone is not always available and can be inaccurate.
Hence we exploit (\ref{equ:rtt_delay}) to avoid synchronization issues. However,
OS delays also play an important part in timestamping. Thus, timestamps are not 
accurate enough due to the OS delay and it is impossible to acquire the exact 
instance of an event with the desired precision.

In acoustic ranging, finding the value of $T_1,T_1',T_2$ and $T_2'$ using the
Android API is nearly impossible because there are unpredictable delays between
the call for \texttt{play()} and the actual played sound (the same holds for the
time between the actual recorded sound and the call of the 
\texttt{startRecording()} method). One can make this time difference smaller 
with some programming tricks, for example running the recording or playing 
procedure on a different thread than the main UI. E.g., there is a priority 
option for the threads called \texttt{AUDIO\_\_PRIORITY} that lets us have low 
latency audio. But even using this approach, there is still a significant time 
delay. The objects for recording and playing in Android are called \texttt{
AudioRecord} and \texttt{AudioTrack} respectively. Both of them have callbacks 
that can be triggered when a certain audio sample is played or recorded. 
Theoretically, one can timestamp the real playing and recording time using these 
callbacks. In practice, the callback itself causes a time delay. We measured up 
to 100 ms extra delay using the callbacks. This time delay has less variations 
compared to the other sources of delay and it seems to be the same for the 
different phones.

To avoid the aforementioned difficulties, we decided to use sample 
counting~\cite{peng2007}. Instead of using timestamps, one can calculate the 
time difference in the recording domain using the number of samples. 
Consequently, each phone records both its own signal and the signal of other 
phones. Hence $\Delta_p$ can be calculated as 
$\Delta_p=(\Delta_1-\Delta_2)/(2f_s)$, where $\Delta_i$ is the number of 
samples between phone $i$'s generated pulse and the other phone's pulse and 
$f_s$ is the sampling frequency (in our case 48 kHz).

In this case the time delay is negligible, even when the size of the recording 
buffer is rather small. However, in this case there is a problem with the 
detection part that will be explained in the next section.

\subsection{Pulse Shape and Detection}\label{sec:ranging_pulse}
In our early experiments, we used a finite duration sinusoid pulse. The
frequency of the pulse can be different for each phone to make 
detection easier. Because we would like to have a non-audible pulse, we choose
18 kHz and 17 kHz for phone 1 and 2 respectively. 
These frequencies are high enough to be hardly audible and are not too high to 
be distorted too much because of the frequency responses of microphones and 
speakers. The length of the pulse is set to 4000 samples at a sampling rate of 48 kHz, thus keeping the duration of the pulse below 0.1 second.

On the detection side we use a matched-filter. If we use the same pulse shape 
for both phones and each phone records both pulses (one from each phone), we
obtain a very large peak and a smaller one in the matched-filter output. The 
former corresponds to the pulse generated by the phone itself and the latter 
corresponds to the received pulse from the other phone. Our experiments indicate
that there is an Automatic Gain Control (AGC) unit in the audio 
recording hardware of the phones, i.e. presence of the bigger peak can affect 
quantization of the Analog to Digital Converter(ADC) and decreases the value of 
the smaller one. We do not have control over the AGC unit but it seems that it 
changes from one recording buffer to another. Hence, we decreased the size of 
the recording buffer to let the AGC adapt itself to the smaller peak.

\begin{figure}
\center	
\includegraphics[scale=0.7]{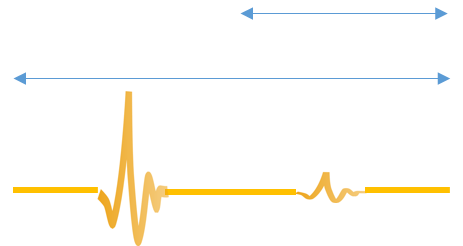}
\caption{We first apply a matched-filter to the whole recorded signal to detect
the largest peak. After finding the larger peak, we truncate the signal and 
apply the filter to the truncated version in order to find the smaller one.
}
\label{fig:twopeaks}
\end{figure}

Because there are two peaks in the recording of each phone, to detect both of 
them, we have to ensure that they are distinguishable. Given this, we can first 
detect the larger peak and then by truncating the recorded signal, we can detect
the smaller peak (Figure \ref{fig:twopeaks}).

With finite duration sinusoid pulse shaping, we can choose different frequencies
to make them more distinguishable. As they have finite duration,
they are not completely orthogonal and can only be distinguished if the
frequency difference is large enough. 
Therefore, we have used pseudo-noise as described in 
Section~\ref{sec:pulse_detection} in later experiments.

\begin{figure}
\center
\includegraphics[scale=0.6]{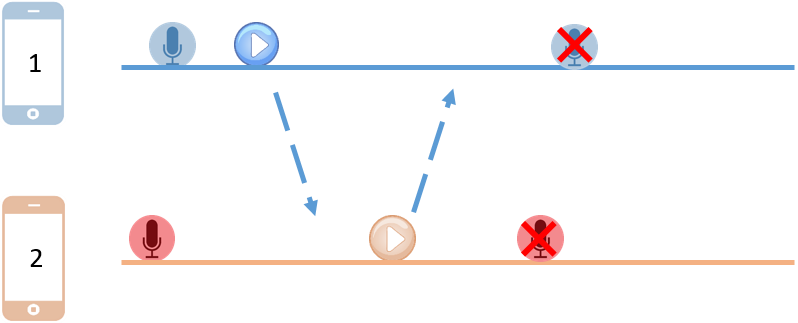}
\caption{Elapsed time between two times of arrivals (ETOA) method: both phones 
start recording (indicated by microphone icon), emit a pulse each (play icon), 
and finally stop recording.}
\label{fig:twophones1}
\end{figure}
\begin{figure}
\center
\includegraphics[scale=0.6]{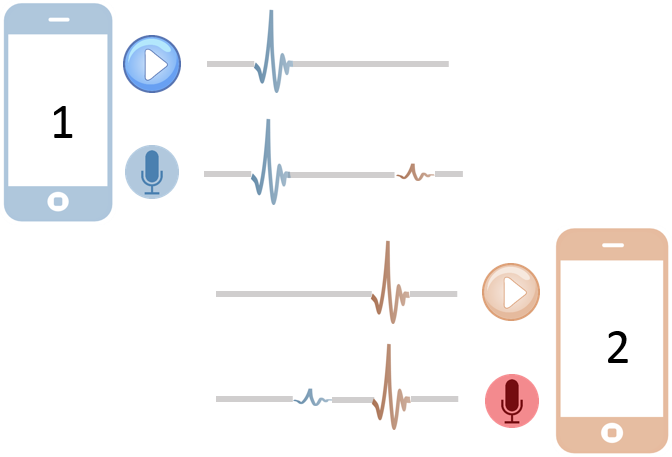}
\caption{Phone 1 sends a pulse, phone 2 responds to it with another pulse, 
everything has been recorded by both phones. The play icon represents what the 
phones are emitting, th microphone icon what the phones are recording.}
\label{fig:twophones2}
\end{figure}

\subsection{Assumptions and Parameter Selection}
The sampling rate $f_s$, recording length and speed of sound $\nu_s$ influence 
the performance of acoustic ranging.

The sampling rate is very important because it determines the maximum frequency 
and bandwidth that can be used. Because the audio hardware of the smartphones 
are designed to play in audible frequency range, the highest possible sampling 
frequency for most of the phones are 48 kHz
~\cite{android_media}. Since according to Nyquist's
theorem a higher sampling rate implies a higher bandwidth, we thus choose 
$f_s=48$ kHz.

The recording length is very important since a short duration can cause 
phones to miss pulses. On the other hand, longer recordings need more memory and
the detection requires more computation. So there is a trade-off between the 
length of the recording and the chance of missing pulses.

To circumvent this issue, we use a simple communication protocol, illustrated in Figure \ref{fig:twophones1}. 
This protocols works over the existing Wi-Fi network. The phones let each other 
know via a WiFi connection that they started recording. After the reception of this message phone 1 
emits its acoustic pulse.
To ensure that phone 1 does not record indefinitely, phone 2 passes a message 
to phone 1 after it played the pulse. As soon as phone 1 receives it, it stops 
recording. This way we are sure that both phones have recorded both pulses and 
no one misses anything. It means that the recording length is not a constant and
it varies according to the OS delays and network delays.

Acoustic distance measurement depends on the speed of sound, which is temperature dependent. Some recent smartphones have temperature sensors. Using this they can 
calculate the speed of sound according to the temperature sensor. As the phones 
we used, Samsung Galaxy S4 Mini, are not equipped with  
such a sensor, we assign the speed of sound according to the average room 
temperature of around $25^{\circ}C$, i.e., $\nu_s=340\,m/s$.

In brief, we built an Android Phone App for acoustic ranging using 
ETOA measurement with sample counting and self-recording to 
calculate the distance between two phones using the above. 
Figures \ref{fig:twophones1} and \ref{fig:twophones2} illustrate the basic 
mechanisms.

\section{Multiple-Node Localization}\label{sec:multi}

Above, we discussed how to measure distances between two devices using 
acoustics. Furthermore we described how Euclidean Distance Matrices (EDMs) can 
be used to infer positions under ideal conditions in Section~\ref{sec:EDM}. We 
now explain how to use these as building blocks to localize
several phones simultaneously under noisy conditions.

First, we describe our method to collect pairwise distances efficiently,
followed by the description of the pulse shape and detection design we used. 
Subsequently, we discuss how to position several devices simultaneously despite 
incomplete and noisy EDMs.

%
%
%
%
%

\subsection{Central Distance Collection}
\label{sec:distance_collection}
We cannot use the application described in Section~\ref{sec:ranging} to measure 
the pairwise distances between several phones as is. With an increasing number 
of phones, several issues arise.

Let $N$ be the number of phones. There are $N\choose 2$ pairwise distances to be
measured. If we measure one distance at a time and each measurement takes $T_m$ 
milliseconds, we need $T_mN(N-1)/2$ in total to do all the measurements.
If the location of some phones change during this time, we get measurements 
which do not correspond to the same positioning of the phones. Therefore, the 
total measurement time should be short as much as possible to guarantee that the
measured values correspond to one configuration of the phones. Otherwise, we 
cannot use the distances to form an EDM. Therefore, instead of doing individual 
pairwise measurements, we propose a scheme to do all the measurements in one 
interval.

Clearly, the number of calculations to be executed by each phone increases as
the number of phones grows. Also there is an extra calculation that we do not 
have in acoustic ranging, namely solving the MDS problem.
Even though this can be done 
in a distributed way efficiently, it requires that 
each phone knows the pairwise distances of all nodes, which requires the
exchange of $O(N^2)$ messages.
Thus we decided to carry out all computations on a server, which incurs a linear
message complexity of $O(N)$ and also reduces battery power consumption in the phones.

In addition, the server is not only used for collecting data and doing the 
calculations, it can also schedule the localization related activities and
minimize the probability of missed pulses or of two pulses of two phone 
colliding.

Here, we summarized the main responsibilities of the \textbf{ server}:
\begin{itemize}
\item Schedule the pulse emitting procedure.
\item Collect recordings from the phones.
\item Calculate distances and EDMs.
\item Run algorithms to solve the MDS problem and localize phones.\\
\end{itemize}

For the measurements, each \textbf{phone} carries out the following steps.

\begin{enumerate}
\item Start recording when receiving \texttt{LISTEN}($t_1,t_2$) command from the server via a WiFi link.
\item Play pulse after time $t_1$.
\item Stop recording after time $t_2$ has elapsed.
\item Send the recording to the server a WiFi link.\\
\end{enumerate}

\begin{figure}
\center
\includegraphics[scale=0.5]{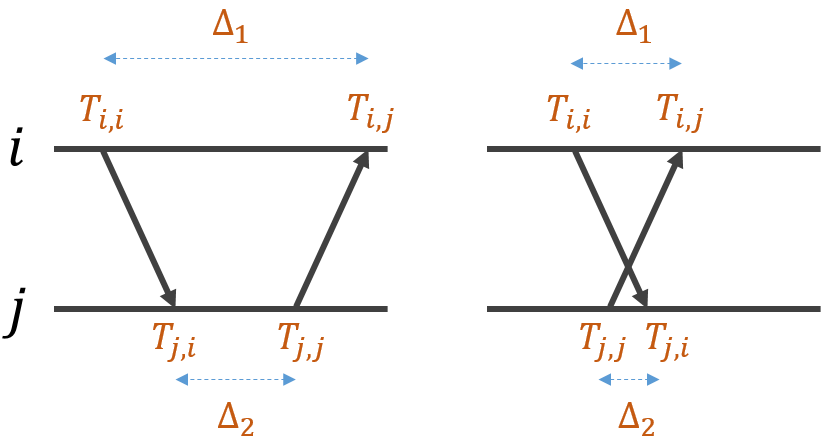}
\caption{(left) RTT $=\frac{\Delta_1-\Delta_2}{2}$. (right) RTT $=\frac{\Delta_1+\Delta_2}{2}$.}
\label{rtt_vector}
\end{figure}

Ideally, in each phone's recording we have $N$ different recorded pulses. If we 
can detect these pulses in each recording, we can find the Round Trip Time (RTT) for each pair. 
The recorded signal of phone $i$ can be written as
\begin{equation}
r_i[n]=\sum_j s_j[n-T_{i,j}],
\end{equation}
where $s_j(.)$ is the received signal from phone $j$ and $n$ is the index of the
$n^{th}$ sample of a signal. $T_{i,j}$ is the index of the sample which 
corresponds to the time phone $i$ receives the pulse from phone $j$. For 
simplicity and because we are only interested in differences, not in absolute 
time, we can assume here that all the phones started the recording at the same 
time. The distance between phone $i$ and $j$ can thus be computed as
\begin{equation}
d_{i,j}=\frac{|(T_{i,j}-T_{i,i})-(T_{j,j}-T_{j,i})|}{2f_s}\nu_s,
\end{equation}
where $\nu_s$ is the speed of sound and $f_s$ is the sampling frequency. Figure 
\ref{rtt_vector} illustrates why this formula is true, 
even if the ordering of pulses leads to negative $\Delta_2$.

\subsubsection{Scheduling: Increasing Reliability}
There are several factors that can cause errors in the measurements, falling in 
one of the two categories.
\begin{itemize}
	\item Android OS and networking delay (missed pulse error).
	\item Acoustics errors (NLOS components, reverberations, obstruction).
\end{itemize}

Consider for example the case where a phone starts recording too late because of
delays introduced by the Android OS and thus misses the pulses of other phones. 
Analogously, a phone can stop recording too early and miss pulses. A good
schedule can minimize the probability of such errors. 

Let $S = \{ (t_1^{(1)}, t_2), (t_1^{(2)}, t_2), \ldots, (t_1^{(N)}, 
t_2)\}$ denote a schedule that tells phone $i$ to emit its pulse $t_1^{(i)}$ 
ms after the reception of the message and to stop recording after $t_2$ ms. The server determines the schedule $S$ and broadcasts it to the $N$ involved phones over the WiFi network. Let $delay_{OS}$ be a bound on the maximum delay cause by the operation 
system and networking. To avoid errors, the schedule computed by the server 
should satisfy some conditions. 
\begin{enumerate}
	\item $\forall i:~ t_1^{(i)} > delay_{OS}$ (to avoid late recording 
	errors)
	\item $\min_i (t_2 - t_1^{(i)}) > delay_{OS}$ (to avoid early stopping
	errors)
	\item $\forall i,j:\:|t_1^{(i)}-t_1^{(j)}|>\delta$ (to avoid colliding 
	pulses)\\
\end{enumerate}

We choose $t_j^{(i)}$s for $N$ phones in the following way
\begin{eqnarray}
t_1^{(i)}&=&D_{delay}+i\cdot D_0.\label{equ:t1}\\
t_2 &=& 2\cdot D_{delay}+N\cdot D_0.
\end{eqnarray}
The reason why we separate $t_1^{(i)}$into two terms is the fact that there are 
two different types of error. The first type is to miss pulses and the second 
one is the collision of pulses. To prevent the former, we force the phones to 
wait for an amount of time, i.e. $D_{delay}$, before the first one sends a pulse
to decrease the probability of missing any pulses because not all phones are in
the recording state yet. We determined experimentally that $D_{delay}=100$ ms is a good choice
taking OS and networking delay into account. Collision errors are avoided by an 
additional amount of delay that varies from phone to phone, i.e. phone $i$ waits
$iD_0$ time before playing its pulse (assuming a pulse duration below $D_0$).
To minimize the collision probability under i.i.d. OS and networking delay, given
measurement time $t_2$, we set $D_0 = (t_2 - 2 D_{delay})/N$. Thus, by 
increasing $t_2$, the recording phase is extended, while the error 
probability is reduced. However, the probability that the 
phones have changed their positions in the meantime increases and higher storage
and computation costs are induced.

Another option to increase the probability of success is to repeat the 
measurements and to combine the results (weighted averages). In particular, one 
can ask the following question.
\begin{quote}
For a given total length of a measurement interval $L_T$, what is the repetition
rate that gives the least possible error?
\end{quote}
In other words, what is the best choice for the number of repetitions $m^*$, 
such that
\begin{equation}
m^*=arg\min_m\mathbb{P}_\text{error}(L)^m
\label{equ:repetition_tradeoff}
\end{equation}
where $\mathbb{P}_\text{error}(L)$ is the probability of error for the length of
measurements $L$ and $L=\lfloor\frac{L_T}{m}\rfloor$.
To this end, we discuss in Section~\ref{sec:IEDM} how to fuse several
(potentially incomplete)
EDMs to get a better accuracy result, i.e. optimum weightings. Given a set of 
assumptions one can thus optimize along the trade-off between the required time 
for the measurements and the accuracy. However, this is out of the scope of this
article. In our evaluation section we show that 5 repetitions provide an 
error margin of around 15cm in a noisy office environment.

Consequently, instead of using $N(N-1)$ pairwise recordings with up to two 
pulses each, we use one recording interval at each device with containing up to 
$N$ pulses. This minimizes time and coordination, enabling evaluation in less 
than $1s$.

\subsection{Pulse Shape and Detection Scheme}
\label{sec:pulse_detection}
The detection scheme described in Section \ref{sec:ranging} for acoustic ranging does not satisfy all
requirements for a multi-phone setting. Since we want to carry out all 
pairwise measurements together, two important issues arise:
\begin{itemize}
\item For $N$ phones, we need $N$ different pulse shapes that are easily 
distinguishable because we would like to find their positions in each recording 
and compute their corresponding $T_{i,j}$s.
\item Each phone receives not only different pulses but also with different
power levels. This means that a recorded signal, contains $N$ different pulses 
where the corresponding power depends on the distance between this phone and the
other ones. The detection scheme should thus not be sensitive to the power 
level.
\end{itemize}

These two issues are related. In theory, if we have orthogonal pulse shapes in 
the recorded signal, we can detect them without too much trouble because
\begin{equation}
\langle s_i[n],s_j[n]\rangle=\delta_{ij}R_j[0]
\label{equ:orthogonality}
\end{equation}
where $s_i(t)$ is the pulse of phone $i$. The output of the matched filter
detector for pulse $j$ is
\begin{equation}
\begin{split}
y_{i,j}[n]&=\sum_{n'} r[n']s_j[n'+n]\\
&=\sum_{n'}\sum_m a_{i,m} s_m[n'-T_{i,j}]s_j[n'+n]\\
&=a_{i,j}R_j[n-T_{i,j}]
\end{split}
\end{equation}
where $R_j[n]$ is the autocorrelation of pulse $j$ and $a_{i,j}$ is the 
amplitude of the received pulse $j$ by phone $i$. The last equality holds 
because of (\ref{equ:orthogonality}). Since the maximum value of the 
autocorrelation function is $R_j[0]$, it is easy to determine $T_{i,j}$ by 
finding the maximum value of the output. In this case, having different power 
levels is no longer a problem.

In practice, there are several factors such as noise and imperfect orthogonality
that make the amplitude of the received signal important for detection. 
In a more general case where the orthogonality does not hold, we get
\begin{equation}
y_{i,j}[n]=a_{i,j}R_j[n-T_{i,j}]+\sum_{m\neq j} a_{i,m}R_{i,m}[n-T_{i,m}],
\end{equation}
where $R_{i,m}[n]$ is the cross-correlation function of pulses $s_i[.]$ and 
$s_m[.]$. We had this issue for two phone distance meter too and we solved it 
using a heuristic approach. However, when the number of phones increases and 
there is no tight coordination between the phones to send pulses (unlike the 
case with only two phones), we cannot use that method because we do not know 
anything about $a_{i,j}$s and how they compare to each other.

One possible solution is to detect the pulses iteratively. We can detect one 
pulse at a time and then cancel its contribution from the recorded signal. We 
repeat this procedure on the canceled recording in the previous step for another
pulse recursively.

For example, suppose there are only two phones. Hence, we have two pulses in 
each recording. To detect these pulses we can first detect the larger one. 
Subtracting this pulse from the recording results in a signal that only contains
one pulse (smaller one). Now we can easily detect the smaller one without 
thinking about the cross-correlation term $R_{1,2}[.]$ because we have already 
removed the larger pulse.

Again, this methods would work perfectly if the pulses were not distorted and 
noise free. In practice, 
we have still a residual of larger pulse after cancellation. 

\subsubsection{Pseudo-Random Binary Sequences}
In principle any pulse shape with a narrow autocorrelation function can be used
in such a localization system. Due to the constraints posed by the
built-in microphone and speaker, we select Pseudo-noise (PN) sequences in the 
frequency range 15-20 kHz and durations of 1000 samples. Though 15 kHz is still 
audible by humans, it is noticed only as a very short pulse.
PN sequences have a large bandwidth with a narrow autocorrelation function. 
These characteristics depend on the length of the sequence and facilitate 
detection. The longer the sequence, the better the detectability.

As the phones receive signals from other phones as well as the one emitted by 
themselves, the signals vary in their power levels. To avoid the problem of 
different power levels if we use the traditional matched filter approach, we 
propose a CDMA-like detection scheme that correlates a binary signal to detect 
pulses. A PN binary pulse shape of length $L$ is defined as
\begin{equation}
s[n]=b_n\quad for\quad n=0,2,\dots,L-1,
\end{equation}
where $b_n$s are realizations of i.i.d. binary random variables with 
$P(b_n=1)=P(b_n=-1)=\frac{1}{2}$. These sequences are suitable as 
they do not convey any information in their amplitude. Hence, additive noise
with reasonable variances can be canceled easily by a sign filter. 
Thus, 
we can ignore the amplitude and apply the matched filter detection on a binary 
sequence. 

\begin{figure}
\center
\includegraphics[scale=0.5]{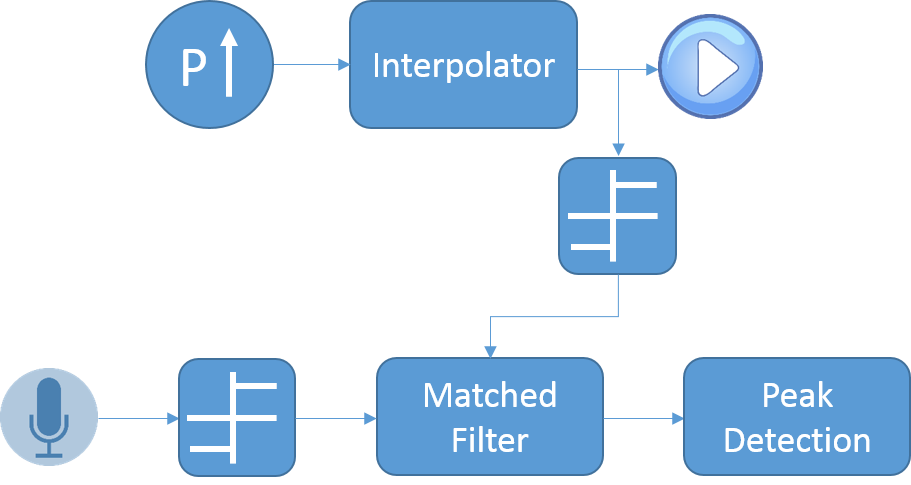}
\caption{Transmitter (top) and receiver (bottom) of detection scheme}
\label{new_detection}
\end{figure}

The proposed detection scheme is illustrated in Figure \ref{new_detection}. On 
the transmitter part, we first upsample the generated PN binary sequence by a 
factor of $P$. For the inserted zeros by upsamplers, we interpolate the values. 
The resulting pulse is our new pulse shape. We do the interpolation and 
upsampling to decrease the required bandwidth and make it lowpass. 
Therefore, we used $P=4$ to reduce the bandwidth and be able to modulate the 
signal to higher frequencies. However, for very high frequencies, greater than 
20 kHz, audio components are more affected by distortions caused by the 
microphone and loud speaker.

On the receiver side, instead of directly applying a matched filter that 
corresponds to the transmitted pulse shape, we pass it through a sign filter. 
Then we apply a matched filter that corresponds to the signed version of the 
pulse shape. The output of the matched filter will be fed into a peak detector 
in order to find $T_{i,j}$s.

The proposed detection scheme shows a better performance compared to using a 
matched filter directly. Though it may be surprising at the first glance, this 
is due to the lack of the optimality condition for the matched filter. The 
matched filter receiver is the optimum linear filter in the sense of SNR. 
However, in this case we do not know the distortion by the acoustic propagation channel, therefore a 
matched filter based on only the pulse shape does not necessarily work better 
in all circumstances.

\subsection{Robust Positioning with incomplete EDMs}
\label{sec:IEDM}
We cannot use the classical method of multi-dimensional scaling discussed in 
Section~\ref{sec:EDM}, as pairwise distance measurements are noisy or
might even be missing if two phones are too far from each other to detect each 
other's pulse. 
There exist optimization-based methods to solve the MDS problem for 
 incomplete EDMs. 
A common cost function to solve this problem is called \textit{raw Stress} 
function~\cite{kruskal1964_2}:
\begin{equation}
\min_{\mathbf{X}\in\mathbb{R}^{N\times\eta}}\sum_{i,j}w_{i,j}
\left(\sqrt{\mathbf{D}(\mathbf{X})[i,j]}-d_{i,j}\right)^2
\end{equation}
This cost function lets us cope with situations where some of the distances are 
not given. We can set $w_{i,j}$ to zero if $d_{i,j}$ is not given. Unfortunately, 
the raw Stress function is non-convex also it is not globally differentiable so 
the optimization methods for solving it are rather involved.

Thus, we use a convex cost function called \textit{s-stress} to solve the 
MDS problem despite incomplete EDMs as proposed by Takane et 
al.~\cite{takane1977}:
\begin{equation}
\min_{\mathbf{X}\in\mathbb{R}^{N\times\eta}}\sum_{i,j}w_{i,j}\left(\mathbf{D}(
\mathbf{X})[i,j]-d_{i,j}^2\right)^2
\label{sstress_cost}
\end{equation}
Contrary to the raw Stress function, the s-stress function is convex and 
differentiable everywhere. 
However, it favors long distances over shorter ones. When $\eta=N-1$, an 
algorithm by Gaffke and Mathar \cite{gaffke1989} can find the global minimum of 
the s-stress function. 
We are interested in cases where the number of spatial dimensions $\eta$ 
($\eta=2$ or $3$) is significantly smaller than the number of nodes $N$. 
Parhizkar \cite{parhizkar2013} proposed an algorithm to minimize this 
function in a (distributed) manner based on the alternating gradient descent 
optimization method, see Algorithm \ref{alg:parhizkar}. In each iteration it 
uses the coordinate descent method by optimizing along one of the variables
cyclically. To the best of our knowledge, no other MDS methods combine 1) 
operation without parameter-tuning, 2) configuration independence, 3) fast 
convergence, and 4) cope with missing/noisy data.
\begin{algorithm}[t]
\caption{Alternating coordinate descent method for minimizing the s-stress 
function \cite{parhizkar2013}}
\textbf{Input:} Distance matrix $\tilde{\mathbf{D}}$.\\
\textbf{Output:} Estimated positions $\hat{\mathbf{X}}$.
\begin{algorithmic}[1]
\State Assume an initial configuration for the sensors $\mathbf{X}_0$.
\Repeat
	\For{sensor number $i=1$ to $N$}
		\State Assume the configuration of the rest of the sensors fixed;
		\State Use the coordinate descent method to find the $x$ coordinate of 
		sensor $i$ using distance information of its neighbors;
		\State Use the coordinate descent method to find the $y$ coordinate of 
		sensor $i$ using distance information of its neighbors;
		\State Send the estimated position of sensor $i$ to its neighbors;
	\EndFor
\Until{convergence or maximum number of iterations is reached.}
\end{algorithmic}
\label{alg:parhizkar}
\end{algorithm}

This algorithm has the advantage that it lets us fuse several sets of 
measurements easily.  Thus, we can increase the accuracy by repeating the 
experiment several times. When there are only two phones, we can simply take the
average over the measured values. Now, suppose we have repeated the measurements
for multiple-node localization and obtained several EDMs, one per measurement. 
The naive approach is to average over each component separately and form a 
new EDM. We can feed this new EDM into Algorithm \ref{alg:parhizkar} to minimize
the cost function in (\ref{sstress_cost}). In other words, this new EDM contains
the mean value of the measured distances. We improve over the naive approach by
weighting the measurements. By modeling the measurement noise  
as additive Gaussian noise, one can show that if we choose 
the weight $w_{i,j}$ inversely proportional to the squared variance of the 
measurements between node $i$ and $j$, the error is minimized, i.e.
\begin{equation}
w_{i,j}=\frac{1/\sigma_{ij}^4}{\sum_{i',j'}1/\sigma_{i',j'}^4}.
\end{equation}
A lower weight reflects a higher variance which means more uncertainty in 
measuring the corresponding distance.
Since we do not know the exact variances of the measurements, we estimate it by
the sample variance.We evaluated both the naive and the optimum weighting 
strategy in Section~\ref{sec:eval_multi}.

\subsection{Application and Server Design}
	The Android application on the phone communicates with the server using a 
	socket connection over Wi-Fi. The first time a phone connects to the server it
	registers itself and goes into waiting mode. 
	Upon a localization request\footnote{Localization requests can be issued by a
	phone, the server or another entity. When and how such requests are 
	triggered is not in the scope of this article},
	the server computes a schedule and broadcasts a separate message with schedule
	information to each registered phones. They start recording as soon as they 
	received the message.
	Each phone plays a generated pseudo-random binary pulse of length $L$ after a 
	certain delay as assigned by the schedule of the server. They stop recording 
	according to the received schedule and send their recorded signal to the 
	server over the wireless link.
	
The server collects all the recordings and processes the collected data to
determine the location of the phones. We implemented our server with Node.JS, 
which is based on Chrome's JavaScript runtime for building fast, scalable 
network applications. It is lightweight, platform-independent and well-suited 
for data-intensive real-time applications. Our application exchanges data in 
JSON format, an open standard format that uses human-readable text to transmit 
data objects consisting of attribute-value pairs.
For the calculation of the position, our Node.JS script calls a server-side 
MATLAB application, allowing for fast prototyping and easy plotting.
\section{Evaluation}
\label{sec:eval}
\subsection{Acoustic Ranging}
\begin{figure}[h]
\center
\includegraphics[scale=0.36]{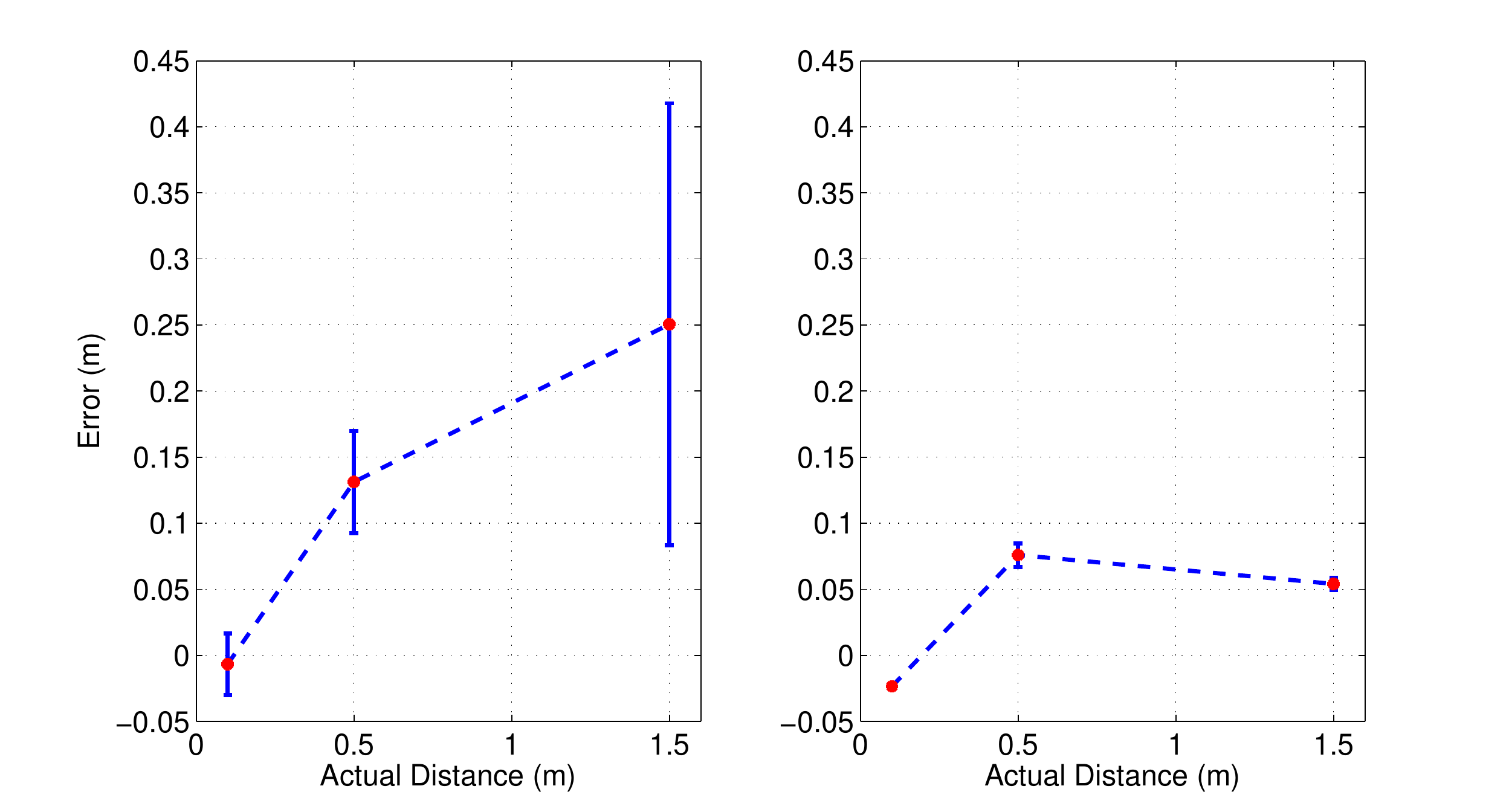}
\caption{Error vs the actual distance for the finite duration 
sinusoid (left) and PN sequence (right) schemes: average and standard deviations of 10 
measurements with phones at distance 0.1m, 0.5m and 1.5m from each other.}
\label{comp_monobin}
\end{figure}

One of the important factors to evaluate an indoor localization system is its 
accuracy. 
Let us first compare the performance of the single tone method described in 
Section~\ref{sec:ranging_pulse} to the CDMA-like approach of 
Section~\ref{sec:pulse_detection}, 
depicted in Figure \ref{comp_monobin}. In the single tone approach, we used two 
sinusoidal pulses at frequencies 18 and 19 kHz for each phone. The accuracy and 
confidence are much better for the binary PN sequence. 

Due to this and the reasons explained in Section~\ref{sec:pulse_detection}, we 
used PN sequences in the remainder of our evaluation. To see how accurate our 
ranging algorithm is, we applied the proposed ranging method with binary PN 
sequences on distances up to 6.5m.
\begin{figure}
\center
\includegraphics[scale=0.34]{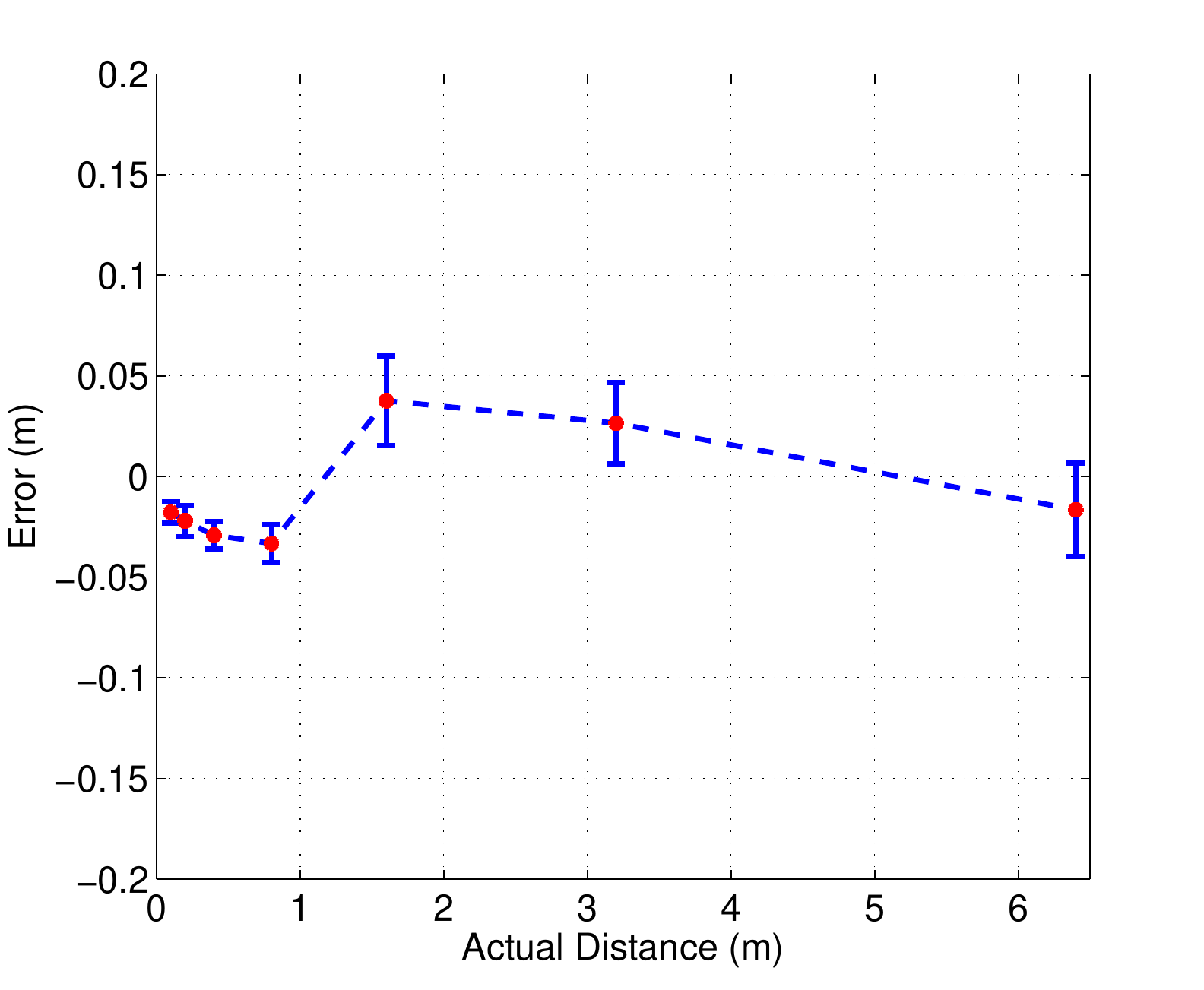}
\caption{Accuracy of the range measurement scheme with PN sequences. 
The error and standard deviation is based on 10 measurements per distance.}
\label{10to640}
\end{figure}
Figure \ref{10to640} shows that the accuracy is around 5cm in this case. Note 
that the data for this plot has been acquired on a different day than the data
for Figure \ref{comp_monobin}, therefore the conditions such as temperature etc. 
differ and also the results are slightly different. As expected, the confidence 
level for longer distances is higher than distances below 1 m. The confidence 
intervals for such accuracy is around 4 cm in the worst case. 
Thus the error for longer distances is surprisingly low, 
i.e., in the order of only 1\%.  
This is due to the fact that we removed distance values exceeding 8m as 
outliers, since they are beyond the effective range of our parameter settings 
for the ETOA approach. Naturally, this has a larger effect when the actual 
distance is larger.


\begin{figure*}[t]
\begin{subfigure}[]{\includegraphics[scale=0.27]{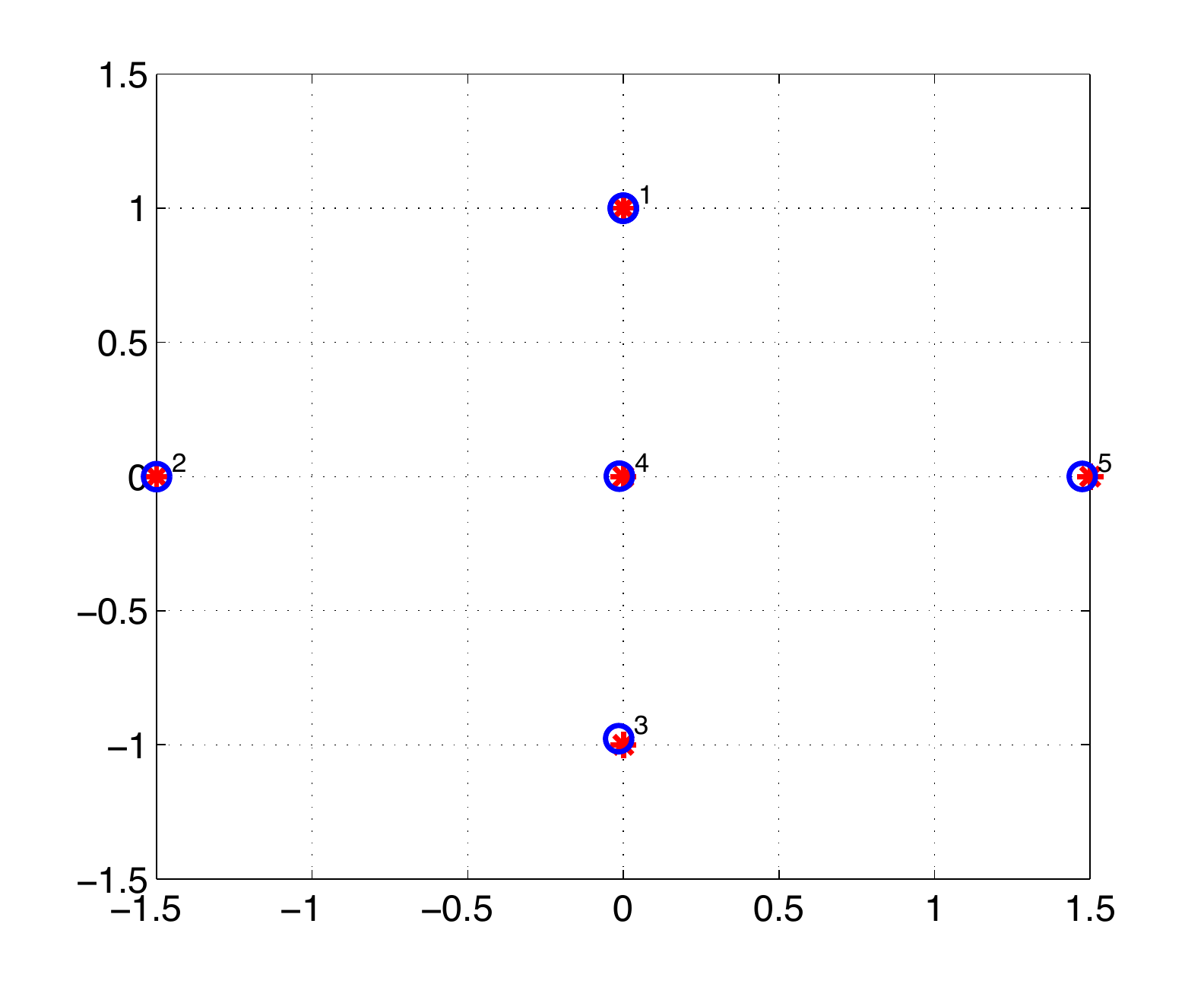}}\end{subfigure}
\begin{subfigure}[]{\includegraphics[scale=0.27]{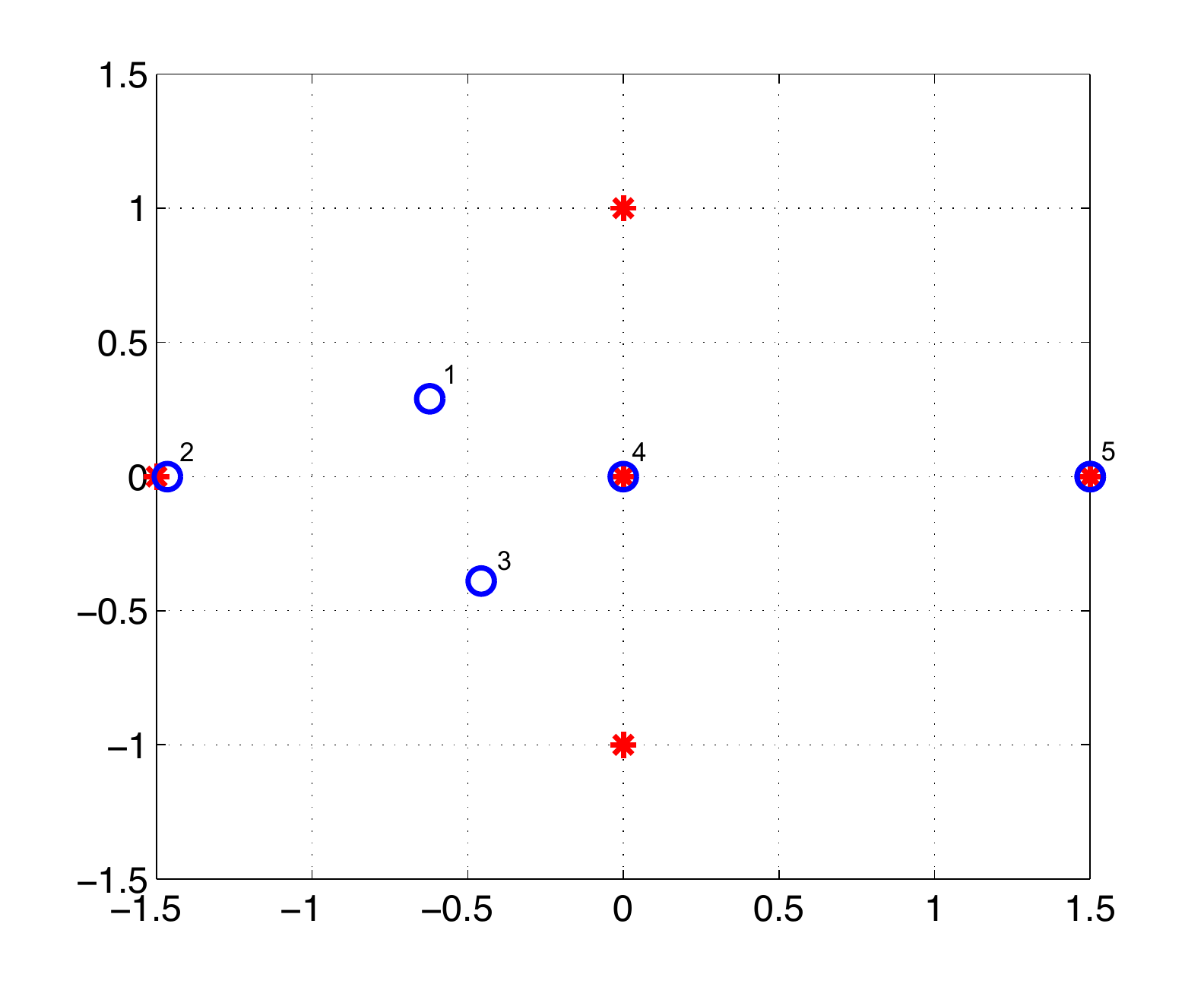}}\end{subfigure}
\begin{subfigure}[]{\includegraphics[scale=0.27]{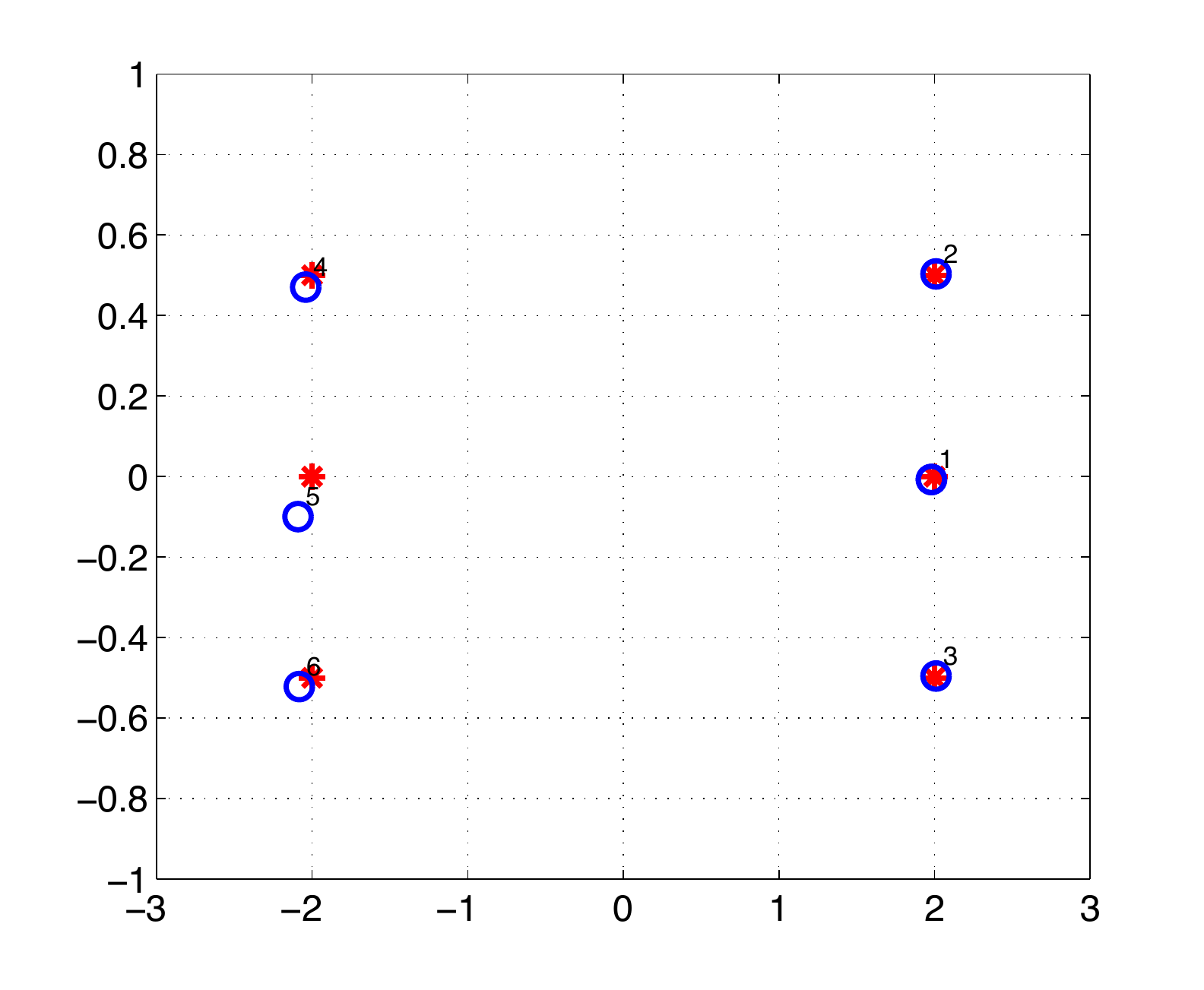}}\end{subfigure}
\begin{subfigure}[]{\includegraphics[scale=0.27]{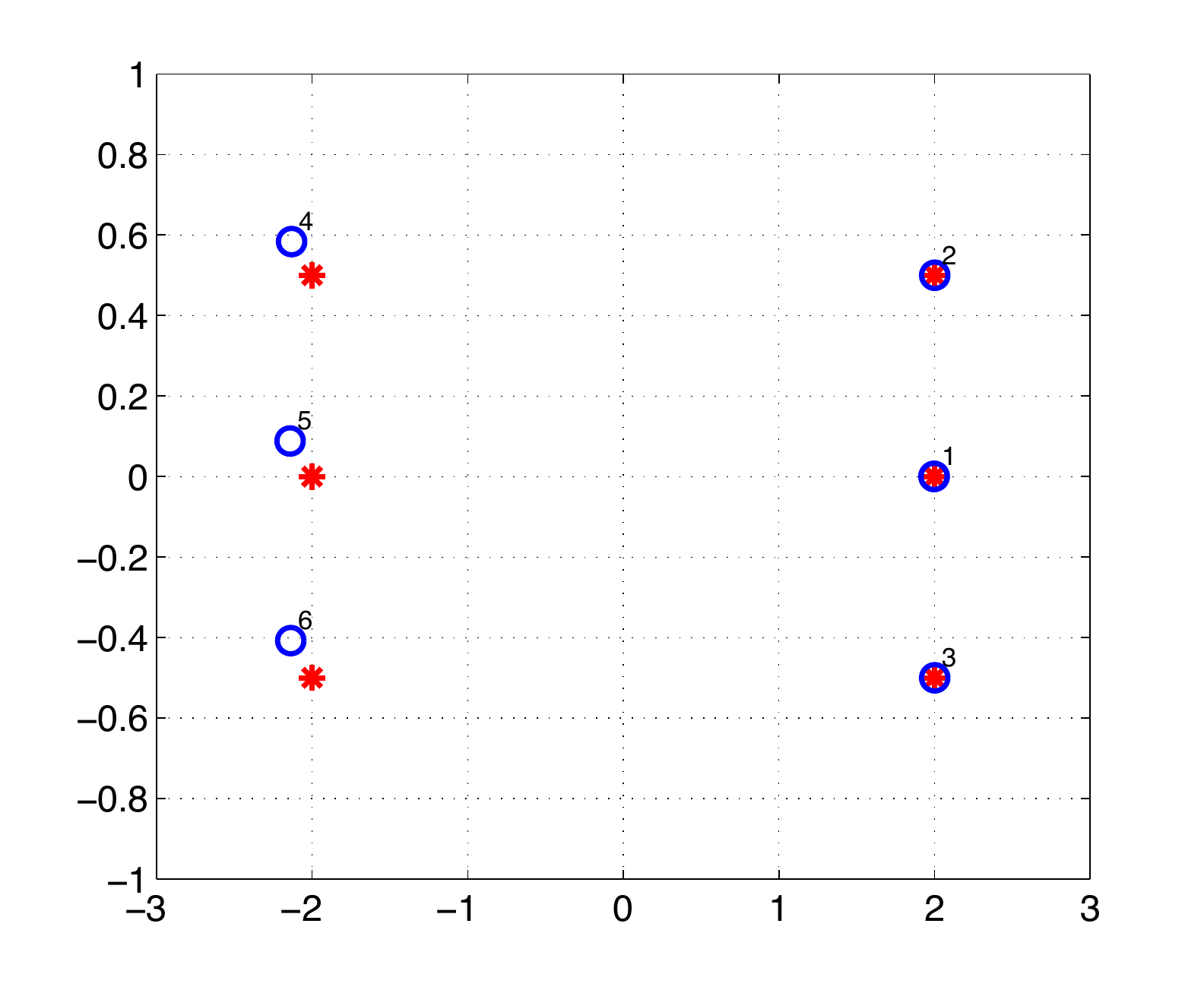}}\end{subfigure}
\vspace{-.1cm}
\caption{Examples with one measurement set. Red 
asterisks correspond to actual,  blue circles to estimated locations. 
(a) and (c) 5 resp. 6 phones in a quiet empty room, 
(b) and (d) 5 resp. phones in an office environment (in meter).}
\label{fig:multiplenode_one}
\end{figure*}

\subsection{Acoustic Multiple-Node Localization}
\label{sec:eval_multi}
Here, we show the results of applying the localization scheme explained in the  
previous section in four different settings. We localized the phones in two 
configurations, a cross-shaped configuration of 5 phones where all the 
distances are mid-range and a three by three configuration of 6 phones where 
the distances are short-range and long-range.
We repeated the experiment in two indoor environments: an empty quiet room.
and an office environment with several people, computers, desks and other 
obstacles and noise.

In Figure \ref{fig:multiplenode_one}, an example of the result obtained from one
set of measurements for each configuration is depicted.
As expected, the accuracy in an office is lower because of the obstacles and 
noise in the environment. It is impossible to keep all influencing factors 
the same in the two environments. For example the quality of the Wi-Fi network, 
which has a great effect on the delays with which the phones start recording,  
varies considerably. The error of the examples in 
Figure~\ref{fig:multiplenode_one} is shown in Table \ref{table_error} in 
centimeter. The error is the average deviation from the actual positions, i.e.
\begin{equation}
e = \frac{1}{N}\sum_{i=1}^N \|\hat{\mathbf{x}}_i-\mathbf{x}_i\|,
\end{equation}
where $\hat{\mathbf{x}}_i$ and $\mathbf{x}_i$ are the estimated and actual 
positions respectively and rotation and translation have been applied for error 
minimization.
The second setup, consisting of 6 phones, shows a better overall performance 
especially in the office. This might be due to the fact that the number of 
phones has a great impact on the performance of the MDS algorithm in \cite{parhizkar2013}. 
However, more experiments are needed to verify this hypothesis. The overall 
accuracy is on the decimeter level.

\begin{table}[h]
\caption{Error comparison (single measurement set).}
\center
\begin{tabular}{|l|l|l|}
\cline{1-3}
 &Cross-shaped &Three-by-Three \\ \cline{1-3}
Empty room &1.30 cm &5.2 cm \\ \cline{1-3}
Office &34.8 cm &8.2 cm \\ \cline{1-3}
\end{tabular}
\label{table_error}
\end{table}

\begin{figure*}[ht]
\center
\subfigure[]{\includegraphics[scale=0.36]{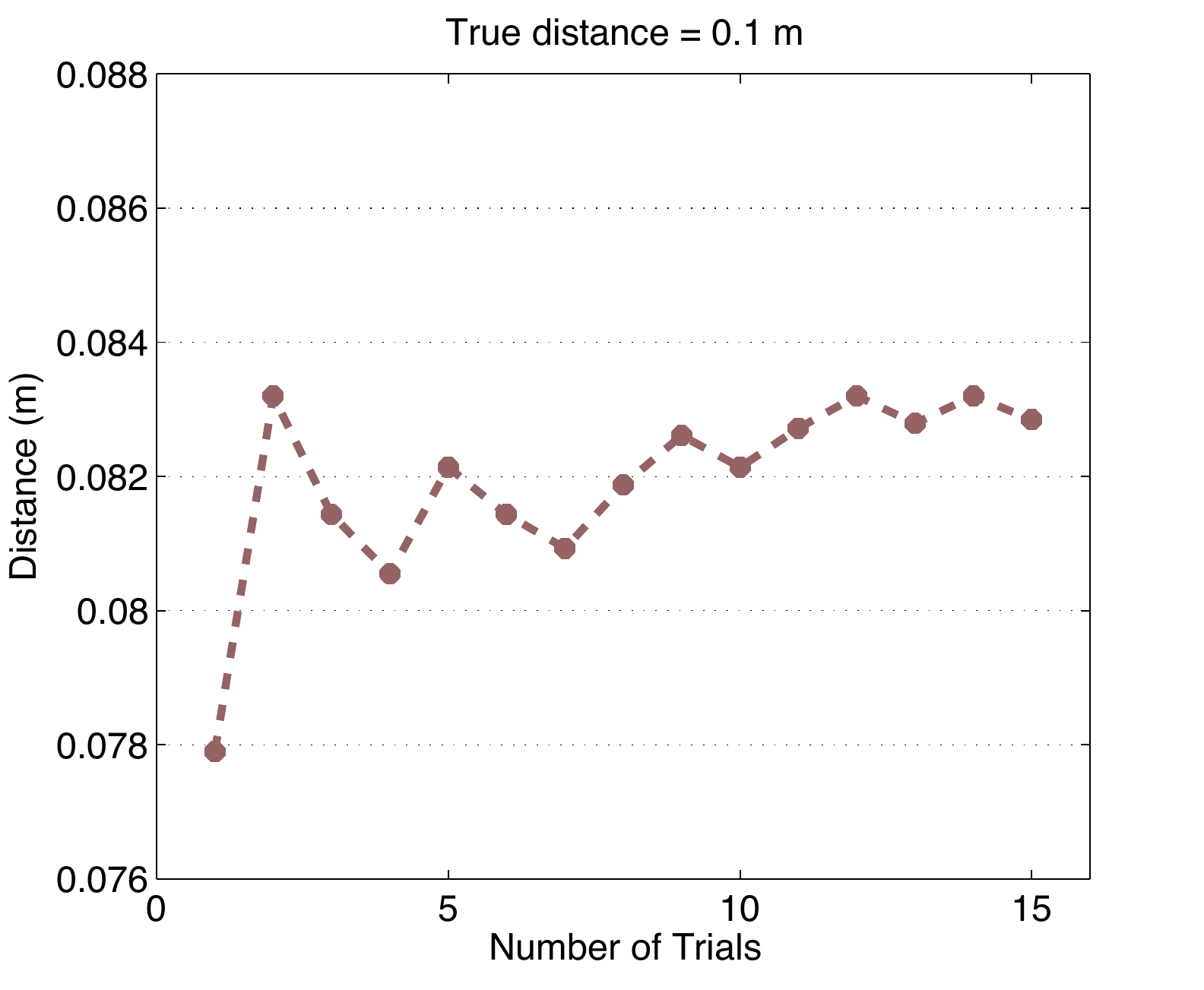}}
\subfigure[]{\includegraphics[scale=0.36]{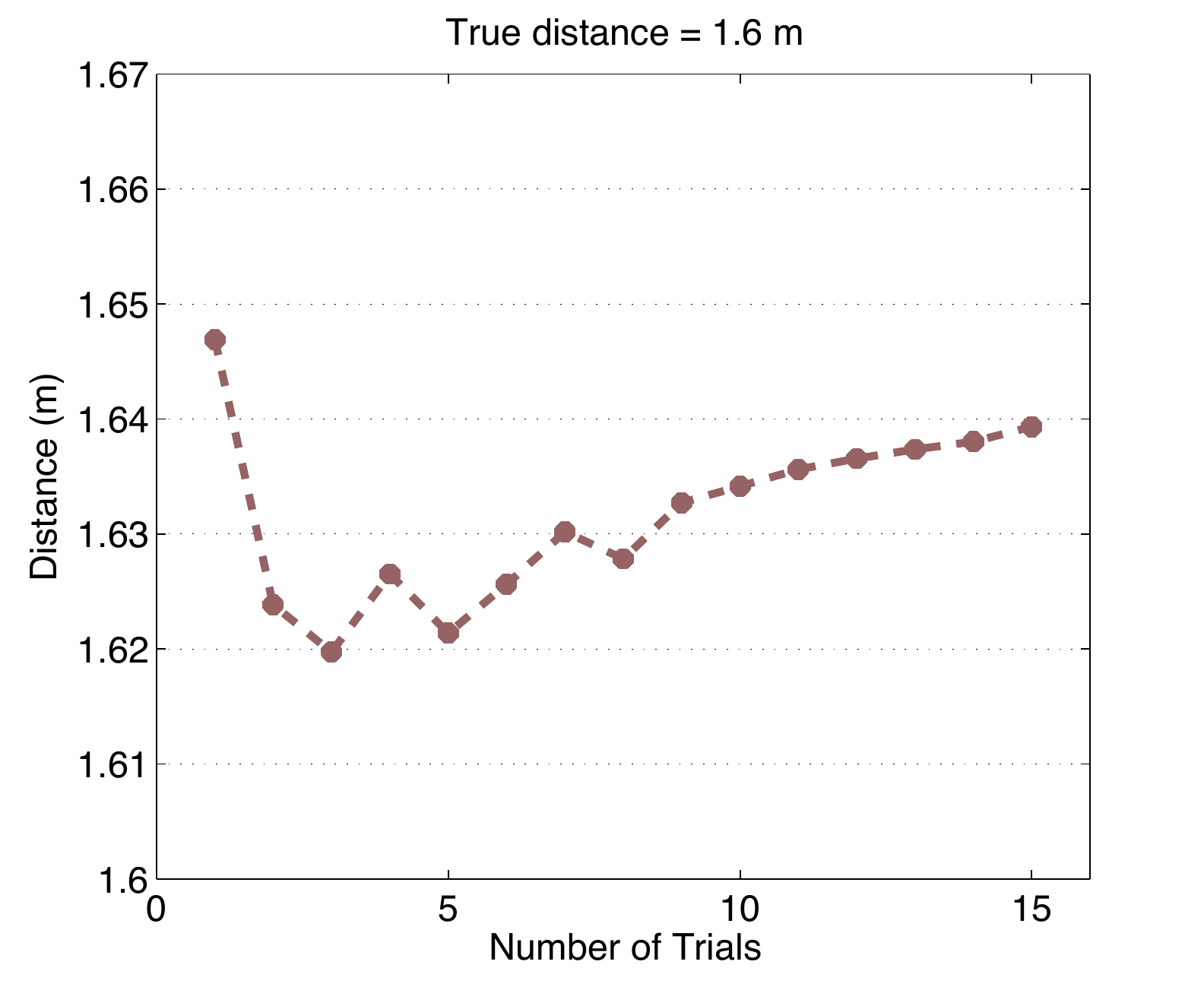}}
\subfigure[]{\includegraphics[scale=0.36]{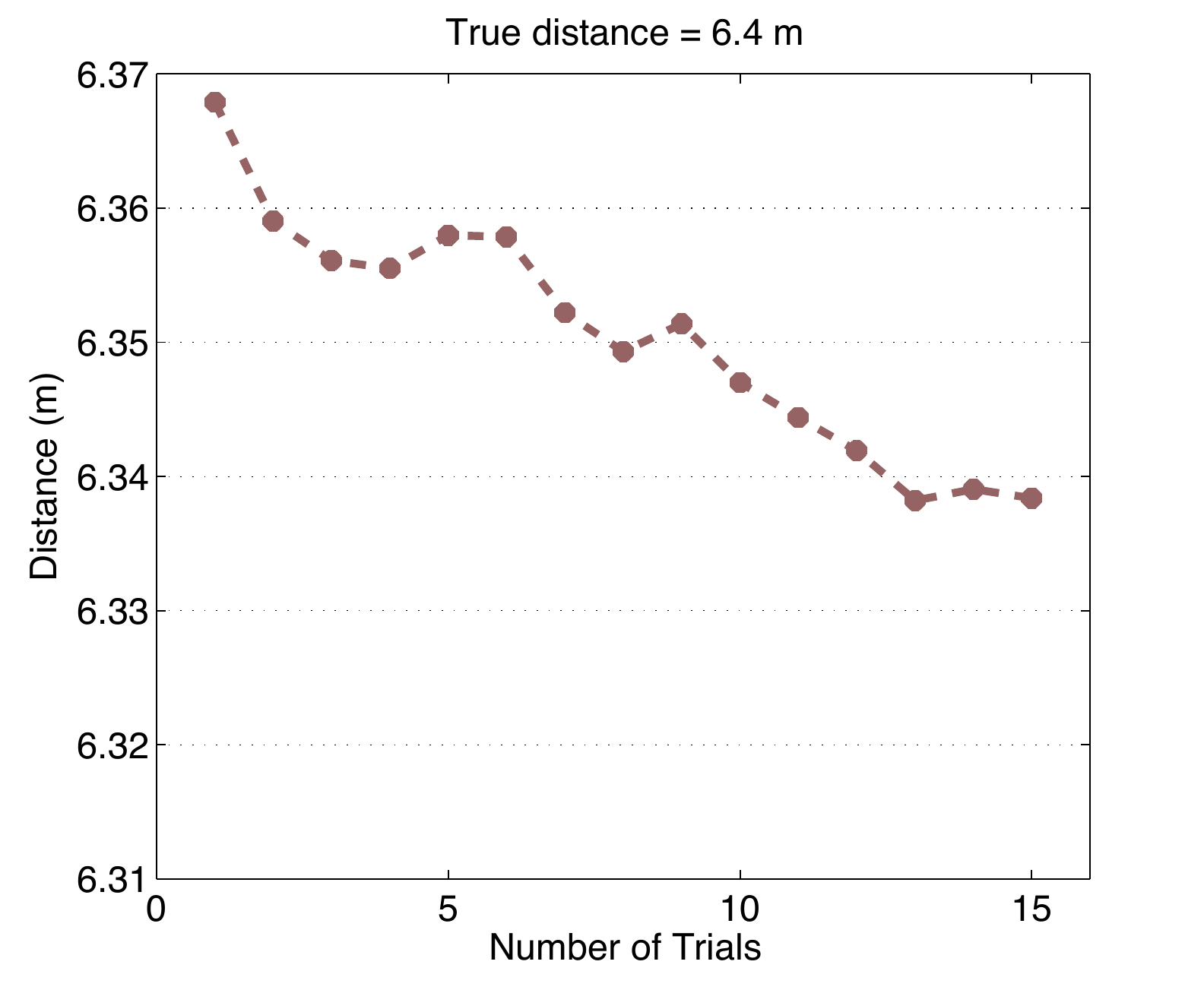}}
\caption{Calculating distances between two phones by averaging over several 
measurements (trials). The filled circles represent the average distance over x 
trials. (a) Distance 10cm:  mean value of 15 measurements is 8.3 cm with 0.5
cm standard deviation. (b) Distance 1.6 m: mean value of 15 measurements is 
1.64 m with 0.019m standard deviation. (c) Distance 6.4 m: mean value of 15 
measurements is 6.34 m with 0.021m standard deviation.}
\label{ranges_trials}
\end{figure*}

To increase the accuracy, several measurements can be carried out and combined, 
as described in Section~\ref{sec:IEDM}.
We evaluate mitigation of the influence of acoustic errors by repeating 
measurements in three different setups: short range, medium range and long 
range. Figure~\ref{ranges_trials}(a) shows the averaged distance versus the 
number of measurements for 10cm distance, while Figures \ref{ranges_trials}(b) 
and \ref{ranges_trials}(c) do the same for 1.6m and 6.4m respectively. 
As expected the short-range setup exhibits lower variance with hardly 
any outlier values. 
Our results indicate that a small number of repetitions, e.g., 5 measurements, 
is sufficient to achieve good accuracy.

\begin{figure*}[ht!]
\center
\begin{subfigure}[Cross-shaped]{\includegraphics[scale=0.4]{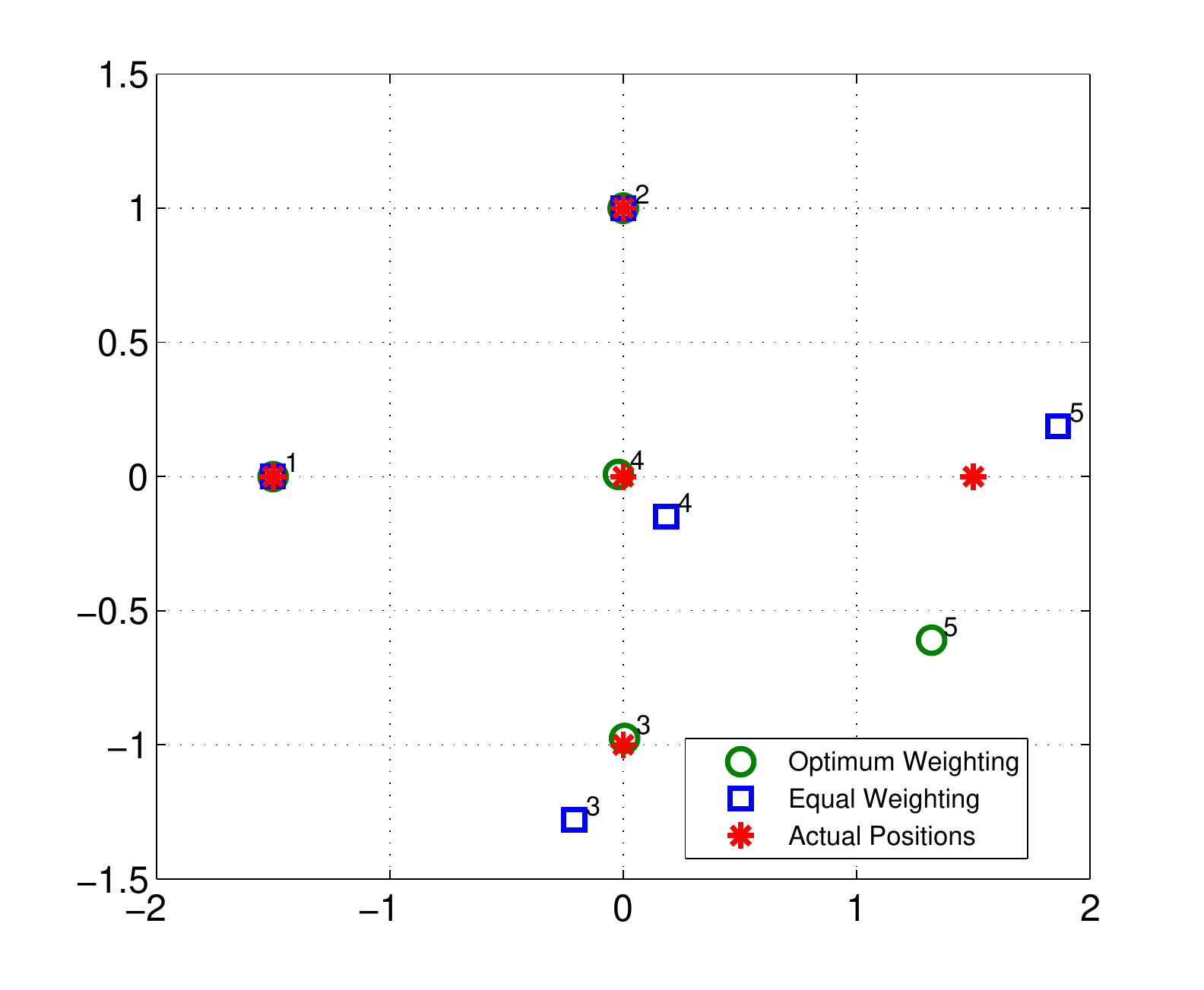}}
\end{subfigure}
\begin{subfigure}[Three-by-Three]{\includegraphics[scale=0.4]{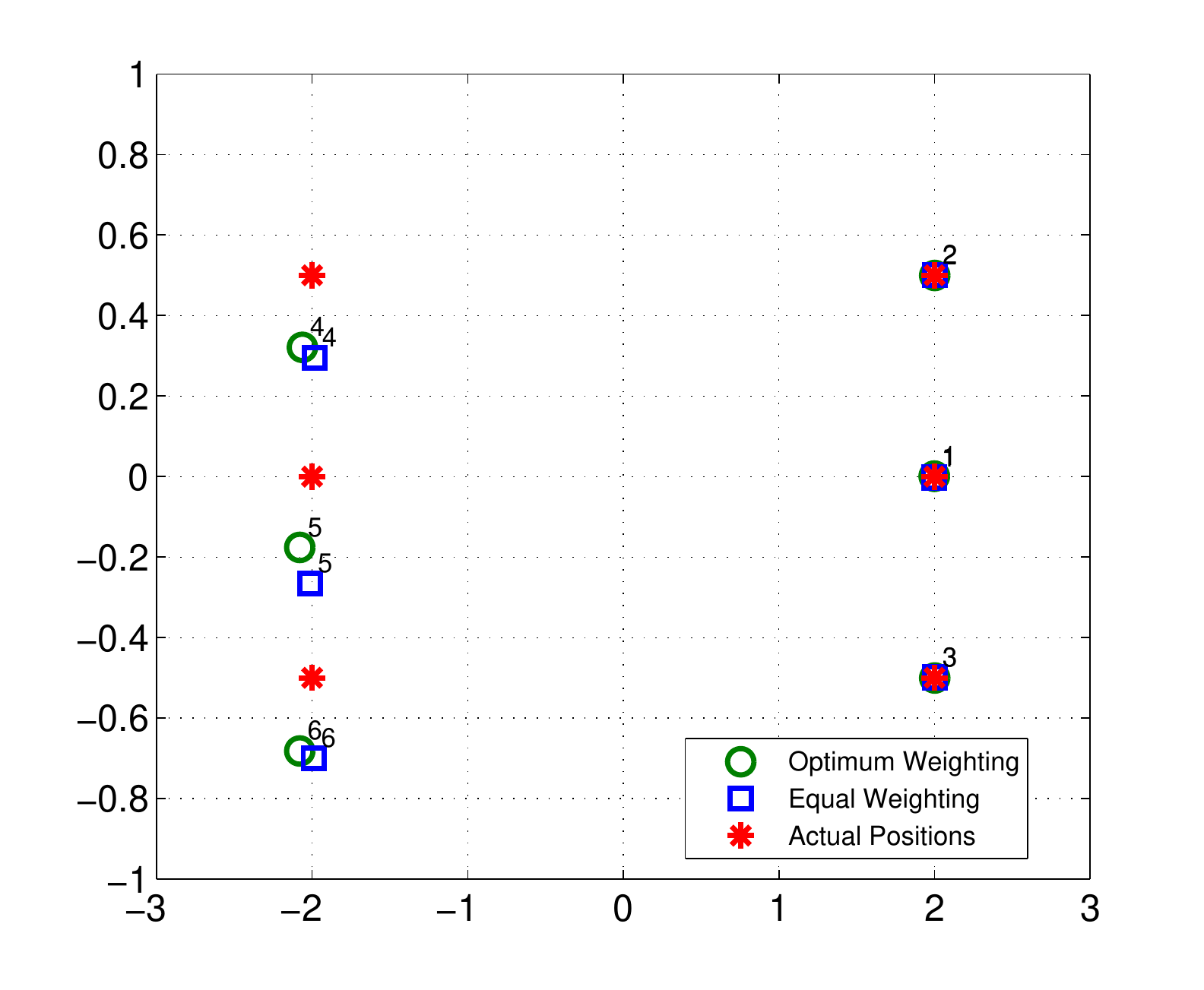}}
\end{subfigure}
\caption{Example of localization results using the optimum weighting scheme for 
5 measurement sets in noisy office.}
\label{opt_fig}
\end{figure*}

\begin{table}[h]
\center
\caption{Comparison of error for different weightings in noisy office (5 sets of
measurements).}
\begin{tabular}{|l|l|l|}
\cline{1-3}
 &Equal Weighting &Optimum Weighting \\ \cline{1-3}
Cross-shaped &19.89 cm &13.6 cm \\ \cline{1-3}
Three-by-Three &11.4 cm &9.7 cm \\ \cline{1-3}
\end{tabular}
\vspace{0.1cm}
\label{table_error_opt}
\end{table}
For multiple-node localization, the effect of repeating measurements  
depends on the weighting strategy. The average distance error is 
shown in Table \ref{table_error_opt}. Compared to the results using one set of 
measurements we observe that error can be reduced up to 50\%. The optimal
weighting scheme outperforms the equal weighting scheme by up to 30\%.
In summary, our localization scheme behaves as expected and is able to provide 
error bounds of around 30cm in noisy environments using one set of measurement 
or around 15cm when combining several measurements.

\section{Related Work}\label{sec:relwork}
Many different indoor localization systems are available today, based on 
pedestrian dead-reckoning, Wi-Fi or other radio signals, cameras, etc. or a 
combination thereof. We refer to \cite{matuz} for an overview. We focus here on 
describing ranging and positioning systems using acoustics.

The fact that many devices can generate sound from their built-in speaker and 
detect sound with the integrated microphone has been used in a number of
different approaches. There are distance-free localization methods which use audio devices to capture acoustic impulse response as an input to a pattern classification algorithm, e.g. \cite{rossi2013}. Here, we focus on the distance-related methods. 
The sound propagation is slow, compared to the speed of radio signals; thus time
stamping signals is easier. Moreover, received  signals can be analyzed in 
detail and the suppression of multipath signals is easier for acoustic signals 
than for radio signals. This helps to increase the accuracy compared to other 
methods.

In the following we describe existing acoustic ranging and positioning systems. 
While there is a multitude of the former, using different pulse shapes and 
calculation methods, the second has not received the same amount of attention. 
From a system design perspective, it is highly valuable to know how to schedule
distance measurements between several nodes with unknown positions and how to 
process the results to derive positions. To the best of our knowledge, the 
current literature does not address these issues.

A large body of work on ranging has been developed in recent years. Among them
BeepBeep~\cite{peng2007} uses smartphones to emit and receive chirp pulses 
between 2 kHz and 6 kHz. The system needs no additional infrastructure and
uses the Round Trip Time (RTT), ETOA and a matched filter to
measure the distance with an accuracy of around 1cm. In~\cite{akiyama2013}, 
authors used Time-difference-of-arrivals (TDOA) and dual-carrier sinusoid with 
500 Hz frequency shift as a pulse shape to locate a single smartphone. The 
indoor localization system Guoguo~\cite{liu2013} uses doublet pulse between 15 
and 20~kHz and Hadamard codes to identify the location of a smartphone. It uses 
beacons transmitted by special hardware and TOA to find the position. 
Other studies focus more on the system design of a localization framework, e.g.,
Beep~\cite{mandal2005}, TOA-based localization system which finds the 
location of a sound source in 3D using Non-linear Least Square Estimation 
(NLSE). ASSIST~\cite{hoflinger2012} uses devices that receive chirp pulses 
between 18 kHz and 21 kHz transmitted from a smartphone to locate it. 
WALRUS~\cite{borriello2005} uses sound emitted from PDAs/Laptops at a frequency
of 21 kHz to identify their location within a room. Hennecke et al. presented a
method for the acoustic self-localization of nodes in an ad-hoc array of COTS 
smartphones. The smartphones worked in the audible range with a short chirp 
impulse between 5 kHz and 16 kHz. The audio signals were received by the 
smartphone microphone \cite{hennecke2011}.  Liu et al.~\cite{liu2013} proposed 
several approaches to improve pulse transmission and achieved 23 cm accuracy by 
averaging over several measurements.
Marziani et al.~\cite{marziani2012} use an RTT and CDMA-based method, to provide
a distributed architecture on specialized devices to find pairwise distances. 
However, they do not determine the actual positions of the nodes.  Chakraborty 
et al.~\cite{chakraborty2014} presented a TDOA-based localization scheme on 
ZigBee modules to determine the location of a sound source with respect to nodes
at a known position. Their solution provides 60cm accuracy using Least Square 
Estimation (LSE). Another localization system using smartphones is 
Whistle~\cite{yu2010},  applying a ranging method similar to BeepBeep with 
10-20~cm accuracy for localizing a sound source with TDOA. It is implemented on 
Windows phones and a server using the method described in \cite{chan1994} for 
solving non-linear TDOA equations to calculate the position of the source. The 
problem of finding the position of multiple nodes simultaneously using acoustics
is not discussed in any of the work we are aware of.

\section{Conclusions}
\label{sec:conclusion}
In this article we proposed a localization system to position several
phones simultaneously. Our system uses ETOA measurements with sample 
counting to compute distances between phones. 
We used the s-stress cost function to 
formulate the problem of finding the positions from the distances as an 
optimization problem to which we applied an alternating gradient descent 
algorithm. 
Furthermore, we described a pseudo-noise-based pulse shaping and detection
scheme and a method to schedule multi-node measurements reliably despite OS and 
networking delays, which has not been addressed in other work to the best of 
our knowledge. 
In addition to reliability, accuracy is an important performance measure of a
localization system. To improve the accuracy, we take measurements several times
and combine them using an optimal weighting strategy.

While we used a centralized approach to do the computations, 
Algorithm \ref{alg:parhizkar} has the intrinsic capability to be implemented in 
a distributed way. The current scheme cannot be directly implemented on the 
phones in a distributed way because of its (message) complexity, therefore 
devising ways to decrease the complexity is a promising path of research.
A more advanced system could also adapt its measurement strategy dynamically to 
changing network conditions and deal with non-line-of-sight (NLOS) components. 

In summary, we want to stress that acoustics can be a powerful tool for indoor
localization and positioning, especially when it is combined with other 
localization methods based on Wi-Fi or inertial sensors. How to fuse several of 
these approaches with acoustic-based methods is another interesting direction for 
further research.

\balance
\bibliographystyle{abbrv}
\bibliography{bib}  

\begin{thebibliography}{10}

\bibitem{akiyama2013}
T.~Akiyama, M.~Nakamura, M.~Sugimoto, and H.~Hashizume.
\newblock Smart phone localization method using dual-carrier acoustic waves.
\newblock In {\em Indoor Positioning and Indoor Navigation (IPIN)}, 2013.

\bibitem{borriello2005}
G.~Borriello, A.~Liu, T.~Offer, C.~Palistrant, and R.~Sharp.
\newblock Walrus: Wireless acoustic location with room-level resolution using
  ultrasound.
\newblock In {\em Conference on Mobile Systems, Applications, and Services
  (MobiSys)}, 2005.

\bibitem{chakraborty2014}
J.~Chakraborty, G.~Ottoy, M.~Gelaude, J.-P. Goemaere, and L.~De~Strycker.
\newblock In {\em Computer and Information Technology (ICCIT)}.

\bibitem{chan1994}
Y.~Chan and K.~Ho.
\newblock A simple and efficient estimator for hyperbolic location.
\newblock {\em Signal Processing, IEEE Transactions on}, 42(8):1905--1915, Aug
  1994.

\bibitem{cover1967}
T.~Cover and P.~Hart.
\newblock Nearest neighbor pattern classification.
\newblock {\em Information Theory, IEEE Transactions on}, 13(1):21--27, January
  1967.

\bibitem{marziani2012}
C.~De~Marziani, J.~Urena, A.~Hernandez, J.~Garcia, F.~Álvarez, A.~Jimenez,
  M.~Perez, J.~Carrizo, J.~Aparicio, and R.~Alcoleas.
\newblock Simultaneous round-trip time-of-flight measurements with encoded
  acoustic signals.
\newblock {\em Sensors Journal, IEEE}, 12(10):2931--2940, 2012.

\bibitem{filonenko2013}
V.~Filonenko, C.~Cullen, and J.~D. Carswell.
\newblock Indoor positioning for smartphones using asynchronous ultrasound
  trilateration.
\newblock {\em ISPRS International Journal of Geo-Information}, 2(3):598--620,
  2013.

\bibitem{gaffke1989}
N.~Gaffke and R.~Mathar.
\newblock A cyclic projection algorithm via duality.
\newblock {\em Metrika}, 36(1):29--54, 1989.

\bibitem{android_media}
Google.
\newblock Android documentation.
\newblock
  \url{http://developer.android.com/reference/android/media/package-summary.html},
  2014.

\bibitem{haykin2013}
S.~Haykin.
\newblock {\em Digital Communication Systems}.
\newblock Wiley, 2013.

\bibitem{hennecke2011}
M.~Hennecke and G.~Fink.
\newblock Towards acoustic self-localization of ad hoc smartphone arrays.
\newblock In {\em Hands-free Speech Communication and Microphone Arrays
  (HSCMA), 2011 Joint Workshop on}, pages 127--132, May 2011.

\bibitem{hoflinger2012}
F.~Hoflinger, R.~Zhang, J.~Hoppe, A.~Bannoura, L.~Reindl, J.~Wendeberg,
  M.~Buhrer, and C.~Schindelhauer.
\newblock In {\em Indoor Positioning and Indoor Navigation (IPIN)}.

\bibitem{kruskal1964}
J.~Kruskal.
\newblock Multidimensional scaling by optimizing goodness of fit to a nonmetric
  hypothesis.
\newblock {\em Psychometrika}, 29(1):1--27, 1964.

\bibitem{kruskal1964_2}
J.~Kruskal.
\newblock Nonmetric multidimensional scaling: A numerical method.
\newblock {\em Psychometrika}, 29(2):115--129, 1964.

\bibitem{liu2013}
K.~Liu, X.~Liu, L.~Xie, and X.~Li.
\newblock Towards accurate acoustic localization on a smartphone.
\newblock In {\em INFOCOM}, 2013.

\bibitem{mandal2005}
A.~Mandal, C.~Lopes, T.~Givargis, A.~Haghighat, R.~Jurdak, and P.~Baldi.
\newblock Beep: 3d indoor positioning using audible sound.
\newblock In {\em Consumer Communications and Networking Conference}, 2005.

\bibitem{matuz}
R.~Mautz.
\newblock Indoor positioning technologies.
\newblock {\em ETH Zurich, Department of Civil, Environmental and Geomatic
  Engineering, Institute of Geodesy and Photogrammetry}, 2012.

\bibitem{parhizkar2013}
R.~Parhizkar.
\newblock {\em Euclidean distance matrices: Properties, algorithms and
  applications}.
\newblock PhD thesis, {\'E}COLE POLYTECHNIQUE F{\'E}D{\'E}RALE DE LAUSANNE,
  2013.

\bibitem{peng2007}
C.~Peng, G.~Shen, Y.~Zhang, Y.~Li, and K.~Tan.
\newblock Beepbeep: A high accuracy acoustic ranging system using cots mobile
  devices.
\newblock In {\em Conference on Embedded Networked Sensor Systems (Sensys)},
  2007.

\bibitem{rossi2013}
M.~Rossi, J.~Seiter, O.~Amft, S.~Buchmeier, and G.~Tr{\"{o}}ster.
\newblock Roomsense: An indoor positioning system for smartphones using active
  sound probing.
\newblock In {\em Proceedings of the 4th Augmented Human International
  Conference}, Mar. 2013.

\bibitem{schoenberg1935}
I.~J. Schoenberg.
\newblock {Remarks to Maurice Fr\'echet’s article “Sur la d\' efinition
  axiomatique d'une classe d'espaces distances vectoriellement applicable sur
  l'espace de Hilbert”. Annals of Mathematics 36(3)}, 1935.

\bibitem{shang2003}
Y.~Shang, W.~Ruml, Y.~Zhang, and M.~P.~J. Fromherz.
\newblock Localization from mere connectivity.
\newblock In {\em Proceedings of the 4th ACM International Symposium on Mobile
  Ad Hoc Networking \&Amp; Computing}, MobiHoc '03, pages 201--212, New York,
  NY, USA, 2003. ACM.

\bibitem{takane1977}
Y.~Takane, F.~Young, and J.~Leeuw.
\newblock {Nonmetric individual differences multidimensional scaling: An
  alternating least squares method with optimal scaling features}.
\newblock {\em Psychometrika}, 42(1):7--67, March 1977.

\bibitem{vucetic1997}
B.~Vucetic and S.~Glisic.
\newblock {\em Spread Spectrum CDMA Systems for Wireless Communications}.
\newblock 1997.

\bibitem{yu2010}
R.~Yu, B.~Xu, G.~Sun, and Z.~Yang.
\newblock Whistle: synchronization-free tdoa for localization.
\newblock In J.~Beutel, D.~Ganesan, and J.~A. Stankovic, editors, {\em SenSys},
  pages 359--360. ACM, 2010.

\end{thebibliography}
\end{document}